\documentclass[%
 aip,
 amsmath,amssymb,
reprint,%
]{revtex4-1}

\usepackage{graphicx}
\usepackage{dcolumn}
\usepackage{bm}

\usepackage[utf8]{inputenc}
\usepackage[T1]{fontenc}
\usepackage{mathptmx}
\usepackage{lipsum}
\usepackage{etoolbox}
\usepackage{amssymb}
\usepackage{amsmath}
\usepackage{amsthm}
\usepackage{mathptmx}
\usepackage{changes,todonotes}
\usepackage{subfigure}
\usepackage[mathscr]{eucal}
\usepackage[colorlinks,linkcolor=red,urlcolor=blue,citecolor=blue]{hyperref}
\usepackage{color}

\usepackage{nameref}

\newcommand{\eref}[1]{Eq.~(\ref{#1})}
\newcommand{\erefs}[1]{Eqs.~(\ref{#1})}

\def\bea{\begin{eqnarray}}
\def\eea{\end{eqnarray}}
\def\ba{\begin{array}}
\def\ea{\end{array}}

\def\la{\langle}
\def\ra{\rangle}

\definecolor{dgreen}{rgb}{0,0.7,0}

\makeatletter
\def\@email#1#2{%
 \endgroup
 \patchcmd{\titleblock@produce}
  {\frontmatter@RRAPformat}
  {\frontmatter@RRAPformat{\produce@RRAP{*#1\href{mailto:#2}{#2}}}\frontmatter@RRAPformat}
  {}{}
}%
\makeatother
\begin{document}

\preprint{AIP/123-QED}

\title[]{Inertial Dynamics of Run-and-Tumble Particle}
\author{Debraj Dutta}
\email{debraj.dutta@bose.res.in}
 \affiliation{S. N. Bose National Centre for Basic Sciences, Kolkata 700106, India.}
\author{Anupam Kundu}%
 \email{anupam.kundu@icts.res.in}
\affiliation{ 
International Centre for Theoretical Sciences, Tata Institute of Fundamental Research, Bengaluru 560089, India
}%

\author{Urna Basu}
\email{urna@bose.res.in}
\affiliation{S. N. Bose National Centre for Basic Sciences, Kolkata 700106, India.}%


\begin{abstract}
We study the dynamics of a single inertial run-and-tumble particle on a straight line. The motion of this particle is characterized by two intrinsic time-scales, namely, an inertial and an active time-scale. We show that interplay of these two times-scales leads to the emergence of four distinct regimes, characterized by different dynamical behaviour of mean-squared displacement  and survival probability. We analytically compute the position distributions in these regimes when the two time-scales are well separated. We show that in the large-time limit, the distribution has a large deviation form and compute the corresponding large deviation function analytically. We also find the persistence exponents in the different regimes theoretically. All our results are supported with numerical simulations.

\end{abstract}

\maketitle

\begin{quotation}
From bacterial motility at micro-scale to flight of birds at macro-scale---active dynamics is ubiquitous in nature. The wide range of potential applications has spurred a significant theoretical and experimental effort to study and characterize active systems, both at collective and individual levels. Most commonly, overdamped dynamics is used to model individual active particles which suffice for describing dynamics of micro-sized entities. However, for larger particles such as insects, birds and artificial agents like vibrobots, it becomes imperative to include effect of inertia. In this paper we have analytically studied the dynamics of an inertial active particle in one dimension, in the absence of any external force.  We have characterized the different dynamical regimes emerging due to the interplay between inertia and activity of an inertial Run-and-Tumble particle. Our analytical results on position distribution can be experimentally verified using artificial active agents like hexbugs\cite{Dauchot2019}.
\end{quotation}

\section{Introduction}\label{sec:intro}
Active matter comprises a wide variety of naturally occurring systems ranging from bacterial colonies to flock of birds\cite{cavagna2014bird, feinerman2018physics, pavlov2000patterns, mukundarajan2016surface} as well as artificially prepared systems like engineered E. coli bacteria or artificial Janus colloids\cite{yang2012janus, Bechinger2016, walther2013janus, nelson2010microrobots}. In such systems, individual particles perform self-propelled `active' motion by consuming energy from their environment\cite{Bechinger2016}, thereby breaking detailed balance even at the level of a single particle. Most commonly studied models of active particles like Active Brownian particles (ABP)\cite{howse2007self, basu2018active}, Run-and-Tumble particles (RTP)\cite{tailleur2008statistical, Malakar2018} and Active Ornstein-Uhlenbeck particles (AOUP)\cite{bonilla2019active, martin2021statistical} are theoretically described using overdamped Langevin equations, assuming that the velocity relaxation takes place over a time scale much smaller than the observation time scale. This description ignores the inertial effects and is fairly accurate for micron-sized particles for which the typical velocity relaxation time is of the order of a few hundred nanoseconds, while the typical temporal resolution of experiments is usually above microseconds. However, for macro-sized particles, such as those with millimeter-scale dimensions, inertial effects become significant because their velocity relaxation time is of the order of several seconds. Consequently, one needs to consider an underdamped description of the active particle motion to describe the effects of inertia on macro-sized particles.

The effect of inertia on active particle motion has drawn a lot of interest in recent years, both at individual and collective levels. It has been shown that the collective properties of active particles are drastically changed due to the presence of inertia. For example, it alters the nature of motility induced phase transition\cite{mandal2019motility, su2021inertia},  promotes hexatic ordering in homogeneous phases\cite{negro2022hydrodynamic}, hinders crystallization\cite{de2020phase, liao2021inertial}, and in general reduces velocity correlations\cite{marconi2021hydrodynamics, caprini2021collective, caprini2021spatial}. Moreover, it has been shown that the presence of inertia makes an active bath behave more like an equilibrium one\cite{khali2024active}. 

Adding inertia can also alter the statistical properties of individual active particles significantly. It has been shown that inertia influences behaviour of ABPs both in free space as well as in the presence of external potential, affecting properties such as kinetic temperature and swim pressure\cite{patel2023exact, patel2024exact, sprenger2023dynamics}. Moreover, the presence of rotational inertia has been found to change the long-time diffusivity for an ABP\cite{scholz2018inertial,breoni2020active,lisin2022motion}. Similarly, the dynamics of inertial AOUP has been studied recently both in the context of free as well as trapped particles\cite{caprini2021inertial}, and also in systems featuring an underdamped self-propulsion force\cite{sprenger2023dynamics}. Futhermore, the behaviour of inertial AOUP has been investigated in the presence of magnetic field\cite{muhsin2022inertial}, Coulomb friction\cite{antonov2024inertial} and in viscoelastic media\cite{adersh2024inertial}. The AOUP being a linear Gaussian process, its position and velocity fluctuations are also Gaussian, fully characterised by the correlation matrix. On the other hand for non-Gaussian processes like ABP, studies are limited to the computation of a few moments\cite{patel2023exact,patel2024exact}. A comprehensive analytical understanding of the effect of inertia on position and velocity distributions is still lacking.

In this paper, we analytically study the effect of inertia on the position distribution of an active particle in the simplest possible setting --- the inertial run-and-tumble particle (IRTP). Run-and-tumble particle is one of the simplest well-studied model of active overdamped motion, where the particle moves along an internal direction which itself changes stochastically\cite{Malakar2018, tailleur2008statistical}. The inclusion of inertia induces a longer delay for the particle to change their speed, increasing the particle mobility and the persistence of the trajectories. The effect of inertia in the stationary state of IRTP in the presence of a harmonic trap has been studied recently\cite{dutta2024harmonically}. In this work we explore the effect of inertia on the dynamical properties of IRTP on a line. More specifically, we analytically characterize the position distribution of the particle in different time regimes arising from the interplay of the activity and the inertial time scales.

\begin{figure}
    \centering
    \includegraphics[width=8.5cm]{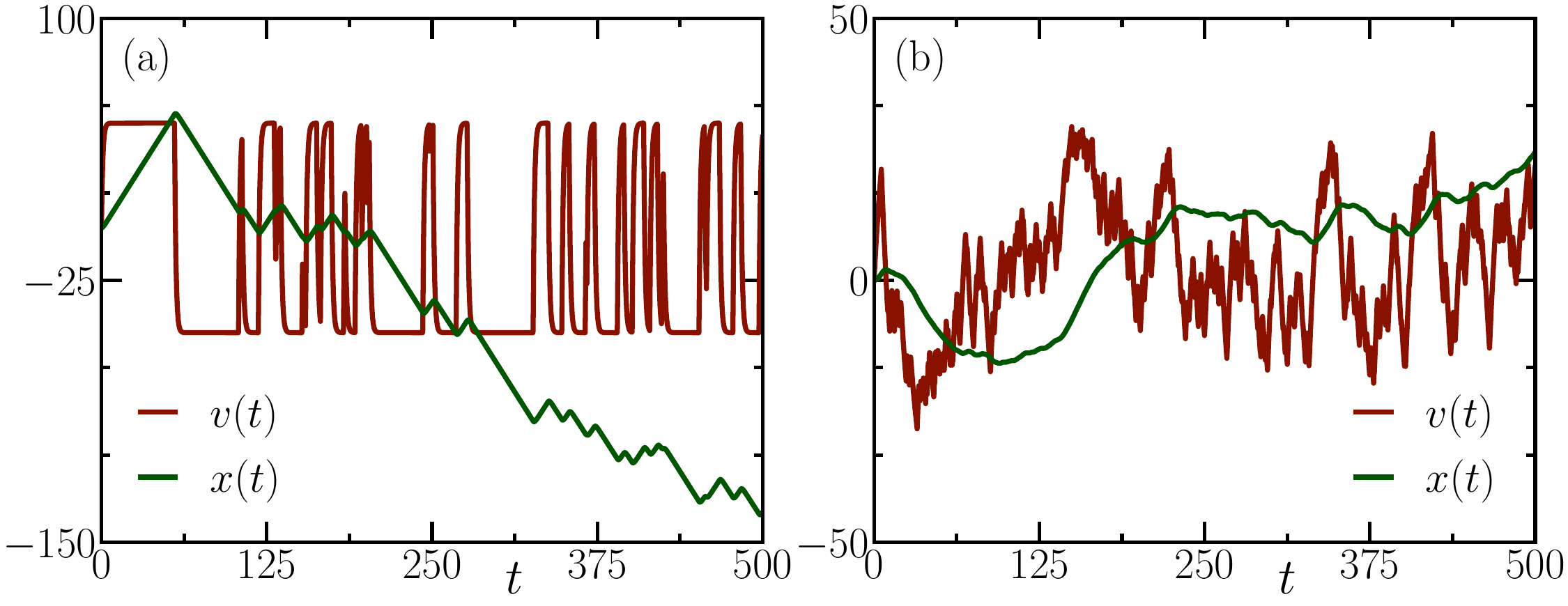}
    \caption{Typical trajectories of the IRTP in the (a) activity-dominated scenario [$\tau_a > \tau_m$] and the (b) inertia-dominated scenario [$\tau_m > \tau_a$]. We have used $\alpha=0.1,m=1.0$ for (a) and $\alpha=1.0,m=0.1$ for (b), while $\gamma=1.0,a_0=1.0$ for both the cases.}
    \label{fig:xvt}
\end{figure}

This paper is organised as follows. In the next section we introduce the inertial Run-and-Tumble particle model and provide a summary of our main results. We then investigate the growth of the position and velocity moments in Sec.~\ref{sec:moments} to get an idea about the dynamical behaviour of the system. In Secs.~\ref{sec:Pvt} and \ref{sec:Pxt} we provide a detailed discussion of the velocity and position distributions in the context of different dynamical regimes arising from the interplay of activity and inertia. A discussion on the first-passage properties and persistent exponents follows in Sec.~\ref{sec:Qt}. Finally, we conclude in Sec.~\ref{sec:conc} with some open questions. The details of the calculations are provided in the Appendix.

\section{Model and Summary of results}\label{sec:model}

We consider the motion of an inertial run-and-tumble particle on a line. The position $x(t)$ and velocity $v(t)$ of the particle evolve according to the underdamped Langevin equations,
\begin{align}
    \dot{x}(t) &= v(t), \label{eq:Le_x}\\
    m\dot{v}(t) &= -\gamma v(t) + a_0 \sigma(t), \label{eq:Le_v}    
\end{align}
where $m$ denotes the mass of the particle and $\gamma$ is the dissipation coefficient. The activity of the particle originates from the self-propulsion force $a_0\sigma(t)$, where $a_0$ is a constant and $\sigma(t)$ is a dichotomous noise that switches between $\pm 1$ at a constant rate $\tau_a^{-1}$. Between two such successive `tumbling' events, when $\sigma$ remains constant, the particle undergoes deterministic damped motion under a constant force $\sigma a_0$. Each tumbling event reverses the direction of this force, alternating between $\pm a_0$. For simplicity, we assume that the particle starts at rest from the origin i.e., $x(0)=0, v(0)=0$. The initial direction of the self-propulsion force is chosen be positive or negative with equal probability $1/2$. 

\begin{figure}
    \centering
    \includegraphics[width=7.5cm]{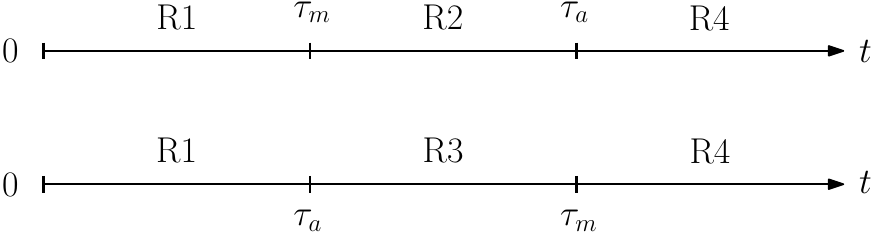}
    \caption{Schematic representation of the different dynamical regimes defined in \eref{regimes}.}
    \label{fig:regimes}
\end{figure}

In the absence of active noise, i.e., for $a_0=0$, the velocity relaxation is controlled by the inertial time-scale $\tau_m=m/\gamma$. Introducing the stochastic active force adds a second characteristic time-scale, $\tau_a$, associated with the particle's activity. Typical trajectories  $\{x(t),~v(t)\}$ of the particle for the cases $\tau_a>\tau_m$ and $\tau_a<\tau_m$ are very different,  as illustrated in Fig.~\ref{fig:xvt}. In fact, depending on the relative magnitude of these time-scales, four distinct dynamical regimes emerge [see Fig.~\ref{fig:regimes}] --- 
\begin{eqnarray}
\begin{split}
        & \text{R1:~Short-time regime -}~&~  t \ll \min(\tau_m,\tau_a) \\
    & \text{R2:~Intermediate-time regime -}~&~ \tau_m \ll t \ll \tau_a \\
    & \text{R3:~Intermediate-time regime -}~&~ \tau_a \ll t \ll \tau_m \\
    & \text{R4:~Long-time regime -}~&~ \max(\tau_m,\tau_a) \ll t, 
    \end{split}
    \label{regimes}
\end{eqnarray}
each characterized by different dynamical behaviour of the particle.  While the velocity dynamics of the IRTP resembles that of an ordinary overdamped RTP in a harmonic trap, a well-studied model, the primary focus of this work is to characterize the behaviour of position fluctuations of the IRTP across these distinct dynamical regimes. Before going to the details of the computation we provide a brief summary of our results.

\begin{itemize}
    \item Using the Fokker-Planck equation, we find a set of recursive relations [see \erefs{eq:ML_1a} and \eqref{eq:ML_1b}] for the general time-dependent moments of the form $\la x^k(t) v^n(t) \ra$. These relations involve specific recursive connections [see Fig.~\ref{fig:mschem} for a schematic representation] that allows computation of the correlations at any order systematically. We find explicit expressions for a few lower order correlations. 
    
    \item From the explicit computations of the mean-squared displacement (MSD)  quoted in \eref{eq:M_20}, we find that it shows distinct behaviour in the different dynamical regimes mentioned above; this is summarized in Table~\ref{table:mom_sct}. 
    
    \item We also characterize the position distribution $P(x,t)$ of the particle in the four different dynamical regimes in \eqref{regimes}. At short-times (in regime R1), similar to overdamped active particles, the IRTP prefers to be drifted away from its initial position as the number of tumbling events is small. This allows us to use a trajectory-based approach to characterize this short-time distribution analytically [see \eref{eq:Pxt_shortt}]. 
    
    \item  At late-times (in regime R4), since the average number of tumbling events is large, the typical position fluctuation becomes Gaussian. However, the atypical fluctuations carry signature of activity and are described by a large deviation function (LDF) [see \eref{eq:LDF_ORTP}], which we compute explicitly. 
    
    \item  The relative strength of activity and inertia becomes crucial in determining the behaviour in the two intermediate regimes R2 and R3. In the activity dominated intermediate regime  R2, the IRTP behaves qualitatively like an overdamped RTP, while in the inertia dominated intermediate regime R3, the IRTP resembles a dichotomous acceleration process without dissipation\cite{santra2023dichotomous}. These mappings allow us to find approximate analytical expressions for the position distribution in these two regimes [see \erefs{eq:Pxt_it_II} and \eqref{eq:Pxt_ir3}]. 
    
    \item The dynamical regimes are also characterized by different persistent exponents. We provide theoretical arguments to predict these exponents which are also verified using numerical simulations. 
\end{itemize}

\begin{table}
\centering
\renewcommand{\arraystretch}{1.5} 
\begin{tabular}{|p{1cm}|p{0.8cm}|p{0.8cm}|p{0.8cm}|p{0.8cm}|}
\hline
               & {~~~R1}      & {~~~R2}      & {~~~R3}      & {~~~R4}  \\ \hline\hline
$~\la x^{2}(t) \ra$ & {\quad$t^{4}$} & {\quad$t^{2}$} & {\quad$t^{3}$} & {\quad$t$} \\ \hline
$~~~\mathscr{Q}(t)$ & ~$O(1)$ & ~~$t^{-\frac 12 }$ & ~~$t^{-\frac 14}$  &  ~~$t^{-\frac 12}$ \\ \hline 
\end{tabular}
\caption{Summary of the leading order behaviour of the MSD and survival probability in the different dynamical regimes.}
\label{table:mom_sct}
\end{table}

\section{Moments of the distribution $P(x,v,t)$}\label{sec:moments}
In this section we provide a systematic method of computing the time dependent moments of $x$ and $v$ for the IRTP. Let $P_{\sigma}(x,v,t)$ denote the joint probability distribution for the particle to have position $x$ and velocity $v$, with active force $a_0\sigma$, at time $t$. The set of Fokker-Planck equations describing the evolution of $P_{\sigma}(x,v,t)$ are given by,
\begin{equation}
    \frac{\partial P_{\sigma}}{\partial t} =-v\frac{\partial P_{\sigma}}{\partial x} - \frac{1}{m}\frac{\partial}{\partial v}\Big[(a_0\sigma-\gamma v)P_{\sigma}\Big]-(P_{\sigma} - P_{-\sigma})/\tau_a, \label{eq:FP_1}
\end{equation}

with $\sigma=\pm 1$. While solving this Fokker-Planck equation directly is challenging, the correlations of the form $\langle x(t)^k v(t)^n \rangle$, for any values of $k$ and $n$, can be efficiently computed recursively using this equation, as we will demonstrate next.
\begin{figure}
    \centering
    \includegraphics[width=6.5cm]{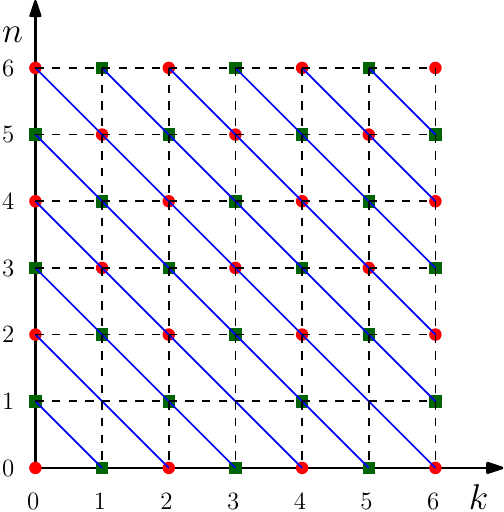}
    \caption{Schematic diagram illustrating the recursive connections between the correlation functions $M(k,n,t)$ (red disks) and $L(k,n,t)$ (green squares) on the $k-n$ lattice.}
    \label{fig:mschem}
\end{figure}

We start by defining the quantities $P(x,v,t)=\sum_{\sigma}P_{\sigma}(x,v,t)$ and $Q(x,v,t)=\sum_{\sigma}\sigma P_{\sigma}(x,v,t)$. Note that, $P(x,v,t)$ is nothing but the joint position-velocity distribution of the IRTP. By recasting the Fokker-Planck equations \eref{eq:FP_1} in terms of the quantities $P$ and $Q$, we get,
\begin{align}
    \frac{\partial P}{\partial t} &= -v \frac{\partial P}{\partial x} - \frac{\partial}{\partial v}\Big[\frac{a_0}{m} Q - \frac{v}{\tau_m} P\Big],\label{eq:FP_2a}\\
    \frac{\partial Q}{\partial t} &= -v \frac{\partial Q}{\partial x} - \frac{\partial}{\partial v}\Big[\frac{a_0}{m} P - \frac{v}{\tau_m} Q\Big] - \frac{2}{\tau_a} Q.\label{eq:FP_2b}
\end{align}
We further define the correlation functions,
\begin{align}
    M(k,n,t) &= \int_{-\infty}^{\infty} dx \int_{-\infty}^{\infty} x^k v^{n} P(x,v,t),\\
    L(k,n,t) &= \int_{-\infty}^{\infty} dx \int_{-\infty}^{\infty} x^k v^{n} Q(x,v,t),
\end{align}
where $k$ and $n$ are non-negative integers. Clearly, $M(k,n,t)\equiv \la x^{k} v^{n} \ra$ refer to the correlations we are interested in. Multiplying both sides of \erefs{eq:FP_2a} and \eqref{eq:FP_2b} with $x^k v^n$, and integrating over $x$ and $v$, we get a set of coupled recursive differential equations,
\begin{align}
    \frac{d}{d t} M(k, n, t) &= k M(k - 1, n + 1) + n \frac{a_0}{m} L(k, n - 1) \cr
    &\hspace{2cm}- \frac{n}{\tau_m} M(k,n,t),\label{eq:ML_1a}\\
    \frac{d}{d t} L(k, n, t) &= k L(k - 1, n + 1) + n\frac{a_0}{m} M(k, n - 1) \cr
    &\hspace{2cm}- \Big(\frac{n}{\tau_m} + \frac{2}{\tau_a}\Big) L(k,n,t).\label{eq:ML_1b}
\end{align}
Note that, to get the above equations we have used the fact that both $P(x, v, t)$ and $Q(x, v, t)$ vanish at $x\to\pm\infty$ and $v\to\pm\infty$. 

By definition, $M(k,n,t)=0$ for negative values of $k<0$ and $n<0$ and the normalization of $P(x,v)$ gives the boundary condition $M(0, 0, t)=1$. We further note the the \erefs{eq:FP_2a} and \eqref{eq:FP_2b} are invariant under the transformations $x\to-x,v\to-v,\sigma\to-\sigma$, which ensures, 
\begin{align}
    M(k,n,t)&=0,\quad \forall (k+n) \in \text{odd},\label{eq:ML_2a} \\
    L(k,n,t)&=0,\quad \forall (k+n) \in \text{even}.\label{eq:ML_2b}
\end{align}
The relations \eqref{eq:ML_1a}-\eqref{eq:ML_2b} along with the normalization condition $M(0,0,t)=1$ allow the computation of the time-dependent correlations $M(k,n,t)$ and $L(k, n, t)$ as discussed below.

Figure~\ref{fig:mschem} shows the $k-n$ lattice where the $d$-th diagonal, given by $k+n=d$, contains $L(k,n, t)$ or $M(k,n, t)$ for odd and even $d$ respectively. We start with the first diagonal $k+n=1$ which contains the entries $L(1,0,t)$ and $L(0,1,t)$. These quantities satisfy the equations,
\begin{align}
    \frac{d}{d t} L(0, 1, t) &= -\Big(\frac{\gamma}{m}+\frac{2}{\tau_a}\Big) L(0,1,t) + \frac{a_0}{m},\label{eq:diff_L10}\\
    \frac{d}{d t} L(1, 0, t) &= -\frac{2}{\tau_a} L(1, 0, t) + L(0, 1, t).\label{eq:diff_L01}
\end{align}
obtained by setting $k=0,n=1$ and $k=1,n=0$ in \eref{eq:ML_2b} along with the boundary condition $M(0,0,1)=1$. The above equations can be solved by first solving for $L(0,1,t)$ using \eref{eq:diff_L10}, and then by substituting $L(0, 1, t)$ into \eref{eq:diff_L01} to solve for $L(0,1,t)$. Using initial conditions $x(t=0)=0$ and $v(t=0)=0$, we explicitly obtain the correlations,
\begin{align}
    L(0,1,t) &= \frac{a_0/\gamma}{1 +2 \tau_m / \tau_a} \Big[1-e^{-t (\frac{1}{\tau_m} +\frac{2}{\tau_a})}\Big],\\
    L(1,0,t) &= \frac{a_0\tau_a/\gamma}{2(1 +2 \tau_m/\tau_a)} \Bigg[e^{-\frac{2t}{\tau_a}} \Big(1 +\frac{2\tau_m}{\tau_a} \big[1-e^{-\frac{t}{\tau_m}}\big]\Big)-1 \Bigg]. \label{eq:L01_L10}
\end{align}
Next we move on the second diagonal $k+n=2$ which contains the entries $M(0,2,t), M(1,1,t)$ and $M(2,0,t)$. Setting $k=0,n=2$, $k=1,n=1$ and $k=2,n=2$ in \eref{eq:ML_1a} we find the equations satisfied by these quantities, which are given by,
\begin{align}
    \frac{d}{d t} M(0,2,t) &= -\frac{2}{\tau_m}M(0,2,t) + \frac{2 a_0}{m} L(0,1,t)\\
    \frac{d}{d t} M(1,1,t) &= M(0, 2, t)-\frac{1}{\tau_m}M(1,1,t) + \frac{a_0}{m} L(1,0,t)\\
    \frac{d}{d t} M(2,0,t) &= 2 M(0, 2, t).
\end{align}

These equations can be solved sequentially, starting from $M(2,0,t)$, subsequently moving down the diagonal solving for $M(1,1,t)$ and then for $M(0,2,t)$. Using the same initial conditions, $x(t=0)=0$ and $v(t=0)=0$, we obtain the correlations appearing on the $d=2$ diagonal, 
\begin{widetext}
\begin{align}
    \la v^2 (t)\ra \equiv M(0,2,t) &= \frac{a_0^2}{\gamma^2} \Bigg[\frac{1}{2 \tau_m/\tau_a  +1}+\frac{2 e^{-t \left(\frac{2}{\tau_a} +\frac{1}{\tau_m  }\right)}}{4 \tau_m  ^2/\tau_a^2-1}+\frac{e^{-2 t/\tau_m }}{1-2 \tau_m/\tau_a  }\Bigg],\label{eq:M_02}\\
    \la x(t)v(t) \ra \equiv M(1,1,t) &= \frac{a_0^2\tau_a}{2\gamma ^2 (2\tau_m/\tau_a-1 )} \Big[\Big(1 -\frac{4\tau_m}{\tau_a}\Big) e^{-t/\tau_m}+ \frac{2\tau_m}{\tau_a} e^{-2t/\tau_m}- e^{-\left(\frac{1}{\tau_m}+\frac{2}{\tau_a}\right)t}+e^{-2 t/\tau_a} + \frac{2\tau_m}{\tau_a}-1 \Big],\label{eq:M_11}\\
    \la x^2 (t) \ra \equiv M(2,0,t) &= \frac{a_0^2 \tau_a^2}{2 \gamma ^2 \left(4 \tau_m  ^2/\tau_a^2-1\right)} \Bigg[\frac{2\tau_m}{\tau_a}   \Big(\frac{2\tau_m}{\tau_a}  +1\Big) \Big(\frac{4\tau_m}{\tau_a}  -1\Big) e^{-t /\tau_m}-\frac{2\tau_m  ^2}{\tau_a^2} \Big(\frac{2\tau_m}{\tau_a}  +1\Big) e^{-2t/\tau_m}\cr
    &+\frac{2\tau_m}{\tau_a}e^{-t\left(\frac{2}{\tau_a}+\frac{1}{\tau_m}\right)}-\Big(\frac{2\tau_m}{\tau_a}+1\Big) e^{2 t/\tau_a}+\Big(\frac{2\tau_m}{\tau_a}  -1\Big) \Big(\frac{2}{\tau_a}\Big[\frac{2  \tau_m}{\tau_a}   t+t-\tau_m \Big(\frac{3\tau_m}{\tau_a}  +2\Big)\Big]-1\Big)\Bigg].\label{eq:M_20}
\end{align}
\end{widetext}
where we have used $L(1,0,t)$ and $L(0,1,t)$ from \eref{eq:L01_L10} obtained in the previous step. One can proceed in a recursive manner to compute the higher-order correlations. In general, the $d$-th diagonal contains $d+1$ entries, satisfying $d+1$ equations. These equations can be solved sequentially starting from $n=d,k=0$ subsequently moving down the diagonal solving for $n=d-1,k=1$, and then for $n=d-2,k=2$ and so on.

\begin{figure*}
    \centering
    \includegraphics[width=17cm]{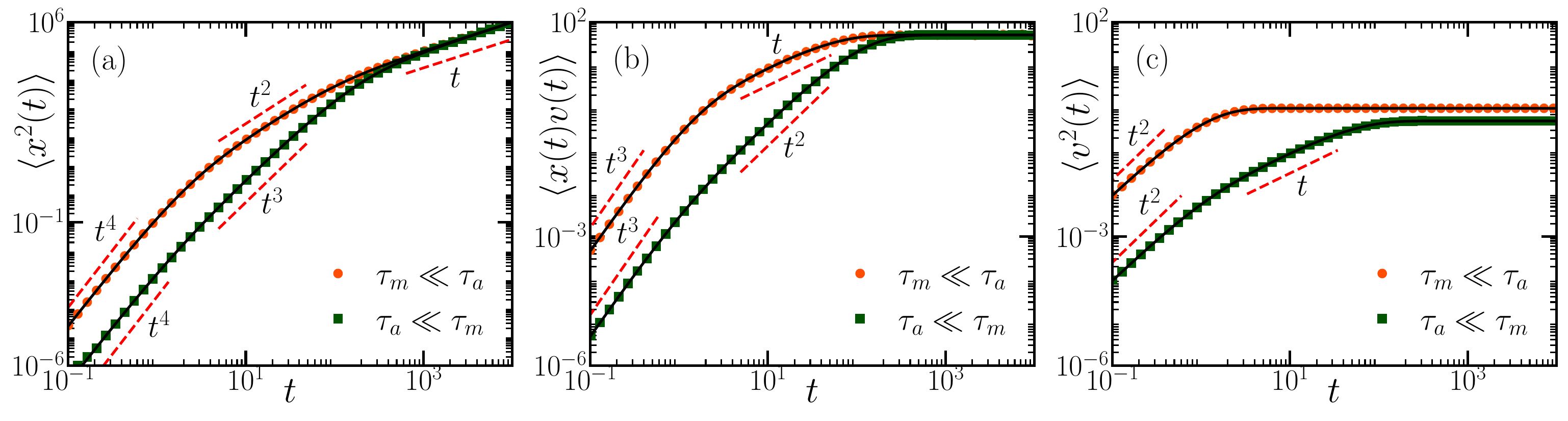}
    \caption{Time evolution of the correlations (a) $\la x^2(t) \ra$, (b) $\la x(t) v(t) \ra$ and (c) $\la v^{2}(t) \ra$ for the cases $\tau_m\ll\tau_a$ and $\tau_a\ll\tau_m$ with $x(0)=0,~v(0)=0$. The symbols indicate the data obtained from numerical simulations while the solid lines indicate analytical predictions [see \erefs{eq:M_02}-\eqref{eq:M_20}]. For the case $\tau_m\ll\tau_a$,  we have used $\tau_a=100.0,\gamma=1.0,m=1.0$, such that $\tau_m=1.0$. For the opposite case $\tau_a\ll\tau_m$, we have used  $\tau_a=1.0,\gamma=0.1,m=10.0$ so that $\tau_m=10.0$. For all the figures we have used $a_0=1.0$.}
    \label{fig:moments_t}
\end{figure*}

It is useful to look at behavior of the mean-squared displacement and the variance of the velocity in the  different dynamical regimes. Since we have assumed that the particle starts at rest from the origin, all odd moments of $x(t)$ and $v(t)$ vanish. Hence, the MSD is nothing but $\la x^2(t) \ra$ and the velocity variance is same as $\la v^2(t) \ra$, quoted in \eref{eq:M_02} and \eref{eq:M_20}, respectively. At short-times, i.e., for $t\ll \min(\tau_m,\tau_a)$, we have,
\begin{align}
    \la v^2(t) \ra = \frac{a_0^2 t^2}{m^2} + O(t^3),\quad
    \la x^2(t) \ra = \frac{a_0^2 t^4}{4m^2} + O(t^5).
\end{align}
In the intermediate regime, this scaling behaviour of the MSD depends on the relative magnitude of the time scales $\tau_m$ and $\tau_a$. For $\tau_m \ll t \ll \tau_a$,
\begin{align}
    \la v^2(t) \ra = \frac{a_0^2}{\gamma^2}+O\left(e^{-t/\tau_m}\right),\quad
    \la x^2(t) \ra = \frac{a_0^2 t^2}{\gamma^2} + O(t^3).
\end{align}
On the other hand for $\tau_a\ll t\ll \tau_m$,
\begin{align}
    \la v^2(t) \ra = \frac{a_0^2\tau_a t}{m^2} + O(t^2),\quad
    \la x^2(t) \ra = \frac{a_0^2 \tau_a t^3}{3 m^2} + O(t^4).
\end{align}
Lastly at long times, i.e., when $t \gg \max(\tau_m,\tau_a) $, the particle becomes diffusive, so that,
\begin{align}
    \la x^2(t) \ra=2 D_\text{eff}  t+O\Big(e^{-t/\min(\tau_m,\tau_a)}\Big),~~ \text{with}~~D_\text{eff} = \frac{a_0^2\tau_a} {2 \gamma^2}.\label{eq:x2_D}
\end{align}
The velocity distribution on the other hand reaches a stationary state at late-times with the variance given by,
\begin{align}
    \la v^{2}(t) \ra = \frac{a_0^2/\gamma^2}{1+2\tau_m/\tau_a}+O\Big(e^{-t/\min(\tau_m,\tau_a)}\Big)
\end{align}
Table~\ref{table:mom_sct} briefly summarizes the leading order behaviour of the MSD in the different dynamical regimes. The time evolution of the second order correlations, $\la v^2(t) \ra$, $\la x(t)v(t) \ra$ and $\la x^2(t) \ra$ are shown in Fig.~\ref{fig:moments_t} which compares results of numerical simulations, obtained using Euler discretization, along with their corresponding analytical expressions \erefs{eq:M_02}-\eqref{eq:M_20}. Note that these variances can as well be computed directly from the Langevin equations \erefs{eq:Le_x}-\eqref{eq:Le_v} and are consistent with the results \erefs{eq:M_02}-\eqref{eq:M_20} obtained using the recursive procedure (see Appendix~\ref{sec:AppB} for details).

\begin{figure}
    \centering
    \includegraphics[width=7.7cm]{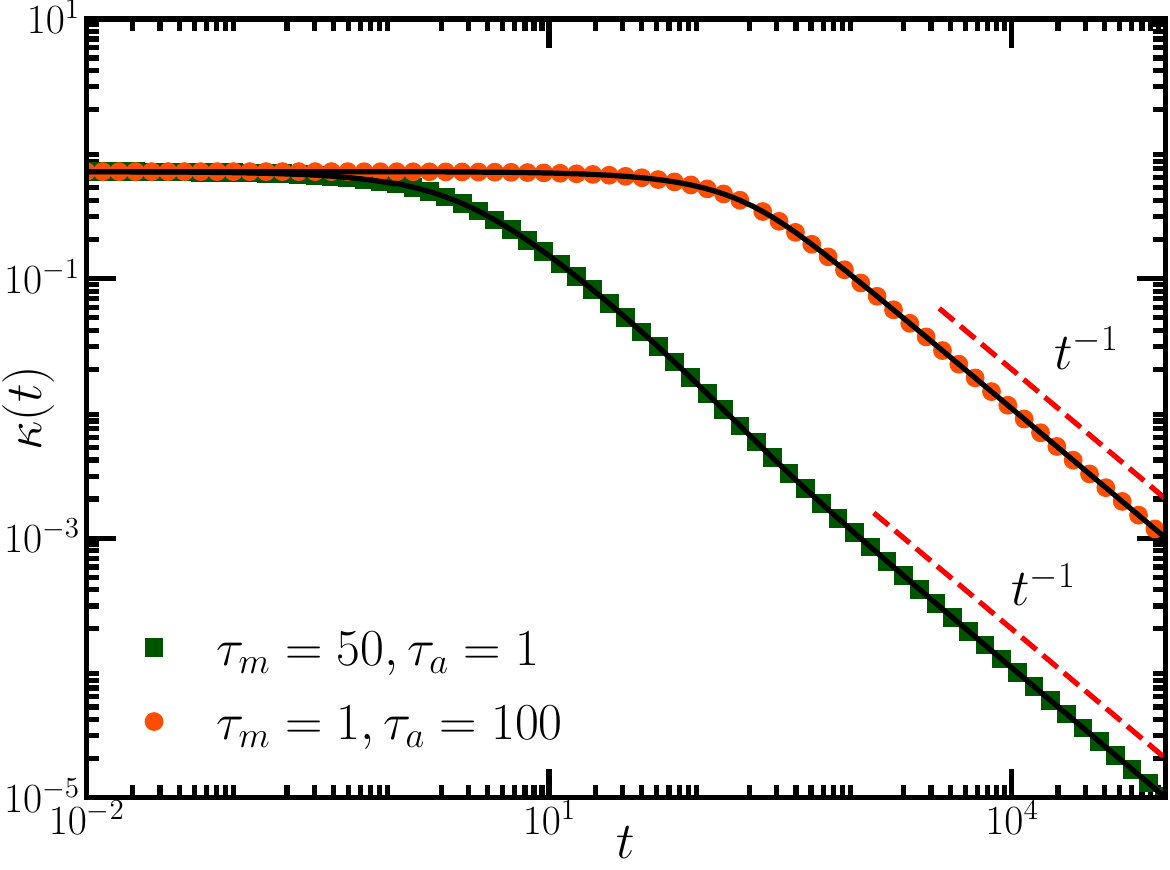}
    \caption{Time evolution of the position kurtosis $\kappa(t)$ [defined in  \eref{eq:kappa}] for the cases $\tau_m\ll\tau_a$ and $\tau_a\ll\tau_m$. The symbols indicate the data obtained from the simulation results whereas the solid black lines indicate analytical predictions [see \erefs{eq:ML_1a}-\eqref{eq:ML_2b}]. Here we have used $\tau_a=100.0,\gamma=1.0,m=1.0$ so that $\tau_m=1.0$ for the case $\tau_m\ll\tau_a$, and $\tau_a=1.0,\gamma=0.1,m=5.0$ so that $\tau_m=50.0$ for the case $\tau_m\ll\tau_a$. For both the cases we have considered $a_0=1$.}
    \label{fig:kurtosis}
\end{figure}

We also compute the kurtosis of the position distribution, defined by,
\begin{align}
    \kappa(t) & = 1-\frac{\la x^4(t) \ra}{3\la x^2(t) \ra^2}, \label{eq:kappa}
\end{align}
using the position moments $\la x^2(t) \ra$ and $\la x^{4}(t) \ra$ obtained by solving the recursion relations to the fourth order. The explicit form of $\kappa(t)$ is rather cumbersome and uninformative. We instead plot the evolution of $\kappa(t)$ in Fig.~\ref{fig:kurtosis} which illustrates its behaviour in the different dynamical regimes. The kurtosis remains constant in the short-time and the intermediate-time regimes, R1, R2 and R3, whereas it shows a $\sim 1/t$ decay at late times (regime R4), indicating eventual convergence of the position distribution to a Gaussian in the $t \to \infty$ limit. The late-time $1/t$ correction to the Gaussian can be obtained from the large deviation analysis, as shown later in Sec.~\ref{subsec:Pxt_lt}.

\section{Velocity distribution}\label{sec:Pvt}
In this section, we provide a brief review of the velocity process of the IRTP which is identical to an ordinary overdamped RTP moving in a harmonic trap, a problem which has been studied previously\cite{dhar2019run,frydel2023run}. For velocity the Langevin equation \eqref{eq:Le_v} when recast in the form,
\begin{align}
\dot{v}(t) &= -\frac{\gamma}{m} v(t) + \frac{a_0}{m} \sigma(t), \label{eq:Le_vx}    
\end{align}
can be thought of as describing the motion of a standard overdamped RTP with self-propulsion speed $a_0/m$ in a harmonic trap of strength $\gamma/m$. At late times, the process has been shown to reach a nonequilibrium stationary state, described by the distribution\cite{dhar2019run},
\begin{align}
    P_\text{st}(v) = \frac{2^{1-2m/(\tau_a \gamma)}}{B\big(\frac{m}{\tau_a \gamma},\frac{m}{\tau_a \gamma}\big)}\frac{\gamma}{a_0}\Big[1-\Big(\frac{\gamma v}{a_0}\Big)^2\Big]^{m/(\tau_a \gamma) - 1},
\end{align}
where $B(u,u)$ denotes the beta function. This distribution has a finite support $ |v| \le a_0/\gamma$ and undergoes a shape transition at a critical value of activity $\tau_a=m/\gamma$. In the active regime $\tau_a>m/\gamma$, the distribution has a U-shape with algebraic divergences occurring at the edges $v=\pm a_0/\gamma$. Conversely, in the passive regime $\tau_a < m/\gamma$, a dome-shaped distribution emerges with $P_\text{st}(v)$ vanishing at the boundaries. Moreover, the relaxation to the stationary state as well as the survival probability have also been studied extensively\cite{dhar2019run}.

\section{Position Distribution}\label{sec:Pxt}

The different dynamical growth of the MSD in different regimes indicates that the position distribution $P(x,t) = \int dv ~ P(x,v,t)$, should also exhibit  distinct behaviour in the different regimes.  In this section, we discuss the behaviour of the position distribution in the four regimes \eqref{regimes} separately. To this end, we recall the initial condition  
\begin{align}
\begin{split}
& x(0)=0,~v(0)=0,  \\
& \sigma(0)= \sigma_1,~\text{with distribution}~ \varsigma(\sigma_1=\pm 1)= \frac 12,
\end{split}
\label{ic}
\end{align}

\subsection{Short time regime (R1)}\label{subsec:Pxt_st}
In the short-time regime, the observation time $t$ is much smaller than both the activity and inertial time-scales $\tau_a$ and $\tau_m$, respectively. Consequently the typical number of changes in the  direction along $\sigma$ is small till time $t$. The trajectory of the particle is piecewise deterministic interspersed by a few tumbling events. In the absence of any tumbling event till time $t$, the particle, starting with the propulsion direction $\sigma_1$, follows the deterministic trajectory,
\begin{align}
    x_{0,\sigma_1}(t)=\frac{a_0 \sigma_1}{\gamma}\Big[t-\tau_m\big(1-e^{-t/\tau_m}\big)],\label{eq:x0t}
\end{align}
obtained by integrating \eref{eq:xt_f}. The corresponding contribution to the position distribution is given by,
\begin{align}
    P_0(x,t)= \frac{1}{2} e^{-t/\tau_a}\Big[\delta(x-x_{0,+})+\delta(x-x_{0,-})\Big],
    \label{P_0(x,t)}
\end{align}
where the factor $e^{-t/\tau_a}$ represents the probability that no tumbling event has occurred till time $t$. 

\begin{figure}[t]
    \centering
    \includegraphics[width=7.5cm]{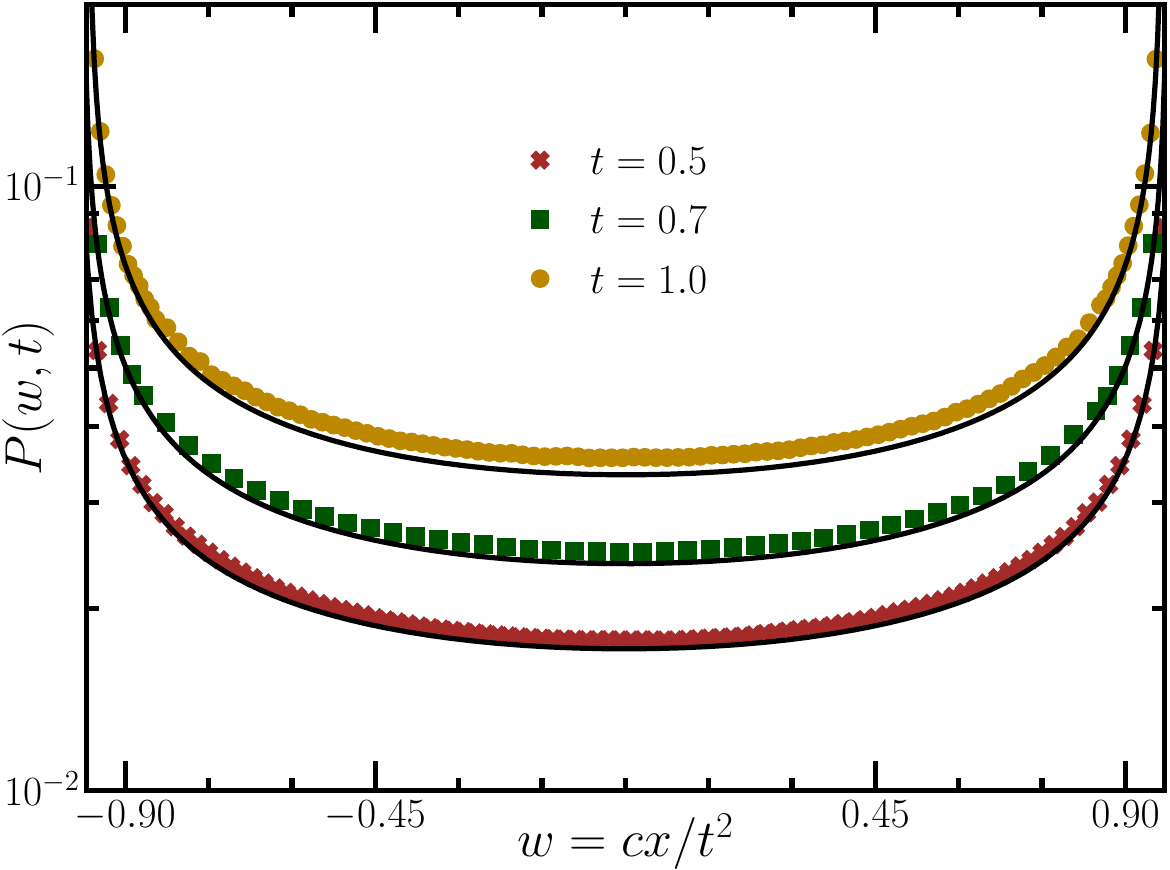}
    \caption{Short-time regime (R1): Distribution of the scaled position $w=cx/t^2$ where $c=2m/a_0$.  Symbols indicate the data obtained from numerical simulations and the solid black lines represent the analytical prediction quoted in \eref{P_0+1(x,t)}. Here we have used $\tau_a=10,\gamma=1.0,m=10.0$ so that $\tau_m=10$ and $a_0=1.0$.
    }
    \label{fig:Pxt_shortt}
\end{figure}

As the contribution from trajectories with non-zero tumbling events are incorporated, the delta-function starts spreading. The leading order correction  $P_{1}(x,t)$ comes from the trajectories with a single tumbling event. To compute this contribution, let us first consider a trajectory where the tumbling event occurs at time $0\le \tau_1 \le t$. Then the particle position $x_{1,\sigma_1}$ after time $t$, starting with $\sigma(0)=\sigma_1$, is given by, 
\begin{align}
x_{1,\sigma_1}(\tau_1) &=\frac{a_0 \sigma_1}{\gamma}\Big[(2\tau_1+\tau_m-t)+\tau_m e^{-t/\tau_m}(1-2e^{\tau_1/\tau_m})\Big].
\end{align}
The probability that, during the interval $[0,t]$ there is a single tumbling event in $[\tau_1, \tau_1+ d\tau_1]$ is given by $[e^{-\tau_1/\tau_a}/\tau_a]d\tau_1 e^{-(t-\tau_1)/\tau_a} = d\tau_1 e^{-t/\tau_a}/\tau_a$. Thus, the contribution to $P(x,t)$ from all possible trajectories with a single tumbling event is given by,
\begin{align}
P_1(x,t)= \frac{1}{2} \sum_{\sigma_1=\pm 1} \left[  \int_0^t d\tau_1~ \frac{e^{-t/\tau_a}}{\tau_a} \delta\Big(x-x_{1,\sigma_1}(\tau_1) \Big)\right].
\end{align}

\begin{figure*}[t]
    \centering
    \includegraphics[width=11cm]{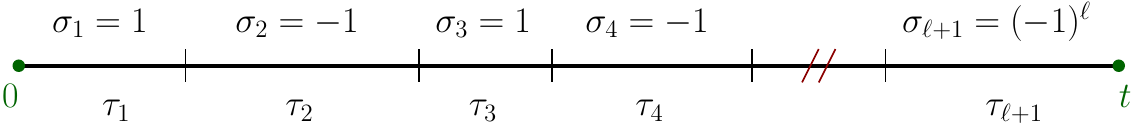}
    \caption{Schematic representation of the tumbling process: $\tau_i$ denotes the interval between the $i^{th}$ and $(i+1)^{th}$ tumbling events during which $\sigma_i=(-1)^{i+1} \sigma_1$ remains constant.}
    \label{fig:s_traj}
\end{figure*}

Performing the integral,  we get,
\begin{align}
    P_{1}(x,t) &=\frac{e^{-t/\tau_a}}{2\tau_a} \sum_{\sigma_1= \pm 1} h_{\sigma_1}(x,t), ~~\text{with}  ~~\label{P_1(x,t)} \\
    h_{\sigma}(x,t) & =\frac{\gamma\sigma}{2a_0}\left[1+W\left(-\exp\left[\frac{1+e^{-\frac t{\tau_m}}}{2}+\frac{t}{2\tau_m} - \frac{\sigma \gamma x}{2a_0\tau_m}\right]\right)\right]^{-1}     \label{eq:Pxt_st_n1}
\end{align}
where $W(\nu)$ denotes the product-log function [see equation (\href{https://dlmf.nist.gov/4.13.E1}{4.13.1}) in Ref.~\cite{NIST:DLMF}]. Adding the contributions $P_0(x,t)$ and $P_1(x,t)$ from trajectories with zero and single tumbling events, respectively from Eqs.~(\ref{P_0(x,t)}) and (\ref{P_1(x,t)}),  already gives a good approximation for the position distribution 
\begin{align}
P(x,t) \approx P_0(x,t) +P_1(x,t), \label{P_0+1(x,t)}
\end{align}
 in the short-time regime. This is illustrated in Fig.~\ref{fig:Pxt_shortt}, which compares this analytical prediction with numerical simulations.
 
 In the following, we generalize this trajectory-based approach\cite{santra2021active} to write the position distribution $P(x,t)$ as a perturbative series in $1/\tau_a$, with the first term corresponding to no tumbling event, the second term corresponding to a single tumbling event and so on.

We consider a trajectory involving $\ell$ flips in $\sigma$ during a time interval $[0,t]$. We can thus divide the duration $[0,t]$ into $\ell+1$ intervals, such that $\sigma$ changes sign at the beginning of each interval and remains constant throughout the interval [see Fig.~\ref{fig:s_traj}]. Let $\tau_i$ be the duration of the $i$-th interval and $\sigma_i=(-1)^{i-1}$ denote the value of $\sigma$ in this interval. We also define $t_i=\sum_{j=1}^{i}\tau_j$ which is the total time elapsed before the start of the $(i+1)$-th interval. Clearly, $t_0=0$ and $t_{\ell+1}=t$. Now, for a given trajectory $\{\sigma_i,\tau_i \}$, the final position $x_{\ell,\sigma_1}$ of the particle starting with propulsion direction $\sigma_1$, can be expressed as a sum of the displacements in each of the intervals $\tau_i$. The detailed calculation is provided in the Appendix~\ref{sec:AppB}. Using the formal solution quoted in \eref{eq:xt_f}, we get,
\begin{align}
x_{\ell,\sigma_1}=(-1)^{\ell}\frac{a_0\sigma_1}{\gamma}\sum_{i=1}^{\ell+1}\int_{t_{i-1}}^{t_i}ds~\big[1-e^{-(t-s)/\tau_m}\big].
\end{align}
To obtain the full position distribution $P(x,t)$ we need to take into account the contributions from all possible trajectories with all possible values of $\ell$. Formally, one can write,
\begin{align}
P(x,t)=\sum_{\ell=0}^{\infty}P_{\ell}(x,t),\label{eq:Pxt_shortt}
\end{align}
where $P_{\ell}(x,t)$ denotes the contribution from the trajectories with $\ell$ number of tumbling events,
\begin{align}
P_{\ell}(x,t)=\frac{e^{-t/\tau_a}}{2\tau_a^{\ell}}\sum_{\sigma_1=\pm1}\int_{0}^{t}\prod_{i=1}^{\ell+1}d\tau_i~\delta\left(t-\sum_{i=1}^{\ell+1}\tau_i\right)\delta(x-x_{\ell,\sigma_1}).\label{eq:Pxtn_shortt}
\end{align}    
The higher order corrections to the position distribution in $\tau_a^{-1}$ can be computed systematically using the above equation. 

Note that, the position distribution in \eref{eq:Pxtn_shortt} is obtained for the particular initial condition in \eref{ic}. However this result
can be  generalized for arbitrary distribution $\varsigma(\sigma_1)$ and initial configuration $(x(0)=x_0,v(0)=v_0)$ in a straightforward manner.

\subsection{Activity dominated intermediate regime (R2)}\label{subsec:Pxt_it_II}
The activity-dominated intermediate regime emerges for $\tau_m\ll t \ll \tau_a$ [see Fig \ref{fig:regimes}]. The limiting scenario $\tau_m\to0$ corresponds to the dynamics of an ordinary overdamped RTP with self-propulsion speed $a_0/\gamma$. This problem is well studied and the position distribution is known exactly \cite{tailleur2008statistical,Malakar2018}, which is given by,
\begin{align}
    P(x,t) &= \frac{e^{-t/\tau_a}  }{2}\Big[\delta\big(x-a_0 t/\gamma\big)+\delta\big(x+a_0 t/\gamma\big)\Big]+\cr
    &\frac{\gamma}{2 a_0\tau_a}e^{-t/\tau_a}\Big[I_{0}(\omega)+\frac{t}{\tau_a\omega}I_{1}(\omega)\Big]\Theta\Big(|x|-\frac{a_0 t}{\gamma}\Big)
\end{align}
where,
\begin{align}
    \omega=\frac{\gamma}{a_0\tau_a}\sqrt{a_0^2 t^2/\gamma^2-x^2}.
\end{align}
Here $\Theta(\nu)$ is the Heaviside step function and $I_n(\nu)$ denotes the $n$-th order modified Bessel function of the first kind. At short times, $t\ll\tau_a$, this distribution has three peaks, which gradually changes into a unimodal shape at late times $t\gg\tau_a$. This transition is characterised by the mean-square displacement showing a crossover from growing linearly with $t$ at short times to quadratically with $t$ at long times.

For an IRTP, the presence of a finite mass is expected to change the position distribution. For small $\tau_m (\ll \tau_a)$ this modification can be computed by deriving an effective equation of motion for the IRTP, as shown below. We start with the formal solution of the Langevin \erefs{eq:Le_x}-\eqref{eq:Le_v}, quoted in \erefs{eq:vt_f}-\eqref{eq:xt_f}. For $x(0)=v(0)=0$, \eref{eq:xt_f} can be recast as,
\begin{align}
x(t)=x_{1}(t)-x_{2}(t),
\end{align}
where the processes $\{x_1(t), x_2(t)\}$ satisfy,
\begin{align}
    \dot{x}_1(t) &= \frac{a_0}{\gamma}\sigma(t),\label{eq:x1_t}\\
    \dot{x}_2(t) &= -\frac{1}{\tau_m}x_{2}(t)+\frac{a_0}{\gamma}\sigma(t).\label{eq:x2_t}
\end{align}
process $x_{1}(t)$ represents a one-dimensional overdamped RTP moving in free space  and $x_2(t)$ represents the same moving in a harmonic trap of strength $1/\tau_m$. Note that, both the processes are driven by the same noise $\sigma(t)$, making them strongly correlated. 

\begin{figure}
    \centering
    \includegraphics[width=7.5cm]{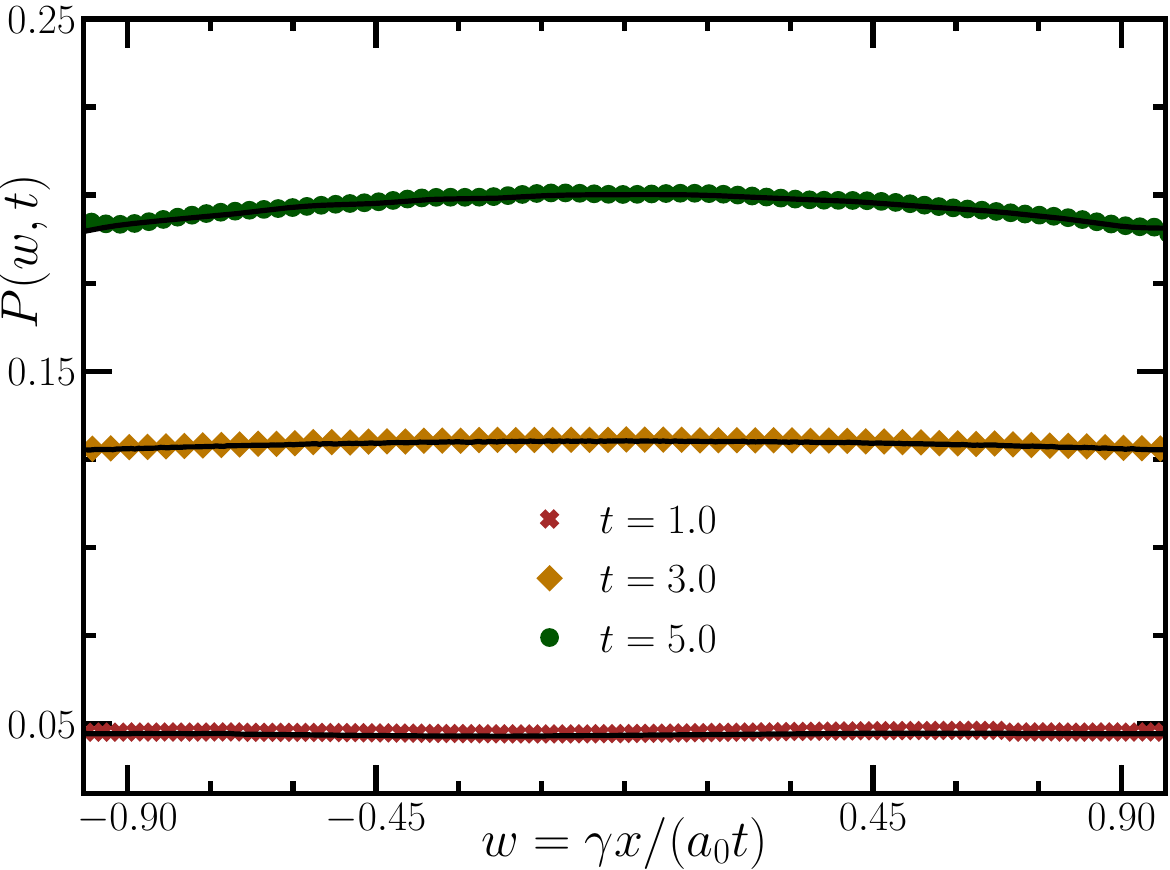}
    \caption{Activity-dominated intermediate-time regime (R2): Distribution of the scaled position $w=\gamma x/(a_0 t)$.  Symbols indicate the data obtained from numerical simulations and solid black lines represent the analytical prediction \eref{eq:Pxt_it_II}. Here we have used $\tau_a=10.0,\gamma=1.0,m=0.01$ and $a_0=1.0$.}
    \label{fig:Pxt_adir}
\end{figure}

Introducing the variable $z(t)\equiv x_{1}(t)+x_{2}(t)$, \erefs{eq:x1_t}-\eqref{eq:x2_t} can be rewritten as,
\begin{align}
    \dot{x}(t) &= \frac{1}{2\tau_m}\big[z(t)-x(t)\big],\label{eq:xt_xz}\\
    \dot{z}(t) &= -\frac{1}{2\tau_m}\big[z(t)-x(t)\big]+\frac{2 a_0}{\gamma}\sigma(t).\label{eq:zt_xz}
\end{align}
Taking time derivative on both sides of \eref{eq:xt_xz} we get,
\begin{align}
\dot{x}(t)=\dot{z}(t)-2\tau_m\ddot{x}(t).\label{eq:xt_expd2}
\end{align}
Substituting \eref{eq:xt_expd2} back in \eref{eq:xt_xz} we get,
\begin{align}
x(t)= z(t)-2\tau_m\dot{z}(t) +4\tau_m^2 \ddot{x}(t).\label{eq:xt_expd3}
\end{align}
Since $\ddot{x}(t)$ is a bounded function, as can be seen from the Langevin \erefs{eq:Le_x}-\eqref{eq:Le_v}, one can neglect the $O(\tau_m^2)$ term for small $\tau_m$. This leads to $x(t) \approx z(t)-2\tau_m\dot{z}(t)$, substituting which in 
 \eref{eq:zt_xz}, we get the following effective equation for $z(t)$,
\begin{align}
\dot{z}(t)=\frac{a_0}{\gamma}\sigma(t)+O(\tau_m^2).\label{eq:zt_expd}
\end{align}
Further substituting \eref{eq:zt_expd} in \eref{eq:xt_expd3} we get,
\begin{align}
x(t)=z(t)-2\frac{\tau_m a_0}{\gamma}\sigma(t),
\end{align}
to the leading order in $\tau_m$. Given the joint distribution $\varrho_{\sigma}(z,t)$ one can compute the position distribution from
\begin{align}
P(x,t) &= \sum_{\sigma=\pm 1}\intop dz~\delta(x-z+2\frac{\tau_m a_0}{\gamma}\sigma)\varrho_{\sigma}(z,t), \label{P(x,t)-R2}
\end{align}
The process $z(t)$ in Eq.~\eqref{eq:zt_expd} is nothing but an ordinary RTP moving freely on a line with propulsion speed $a_0/\gamma$. The characteristic function $\tilde{\varrho}_{\sigma}(q,t)=\int_{-\infty}^{\infty}dz~e^{iqz}\varrho_{\sigma}(z,t)$ of the corresponding joint distribution $\varrho_{\sigma}(z)$ is known exactly\cite{Malakar2018},
\begin{align}
\tilde{\varrho}_{\sigma}(q,t) &= \frac{e^{\varphi_{-}t/\tau_a}}{2}\Bigg[\frac{\big(i a_0\tau_a q-\gamma[1+\varphi_{-}]\big)^2}{\gamma^2\mathcal{N}_{\sigma}^2}+\frac{1}{2(1+\varphi_{-})}\Bigg]\cr
&\hspace{0.05cm}+\frac{e^{\varphi_{+}t/\tau_a}}{2}\Bigg[\frac{1}{\mathcal{N}_{\sigma}^2}+\frac{1}{2(1+\varphi_{+})}\Bigg]\label{tilderho(q,t)}
\end{align}
where,
\begin{align}
    \varphi_{\sigma} &= -1+\sigma\sqrt{1-a_0^2\tau_a^2 q^2/\gamma^2},~~~\text{and},\\
    \mathcal{N}_{\sigma} &= \sqrt{2(1+\varphi_{\sigma})}\Big[(1+\varphi_{\sigma})+i a_0\tau_a q/\gamma\Big]^{1/2}
\end{align}
The above expression can be used in Eq.~\eqref{P(x,t)-R2} to compute  the position distribution $P(x,t)$ of the inertial particle as,
\begin{align}
    P(x,t) &= \sum_{\sigma}\intop_{-\infty}^{\infty}\frac{dq}{2\pi}~\exp\Big[-iq\Big(x+\frac{2\tau_m a_0}{\gamma}\sigma\Big)\Big]\tilde{\varrho}_{\sigma}(q,t). 
\label{eq:Pxt_it_II}
\end{align}
Although the $q$-integral cannot be performed analytically, it can be evaluated numerically to obtain $P(x,t)$ to arbitrary accuracy. This is illustrated in Fig.~\ref{fig:Pxt_adir} which compares the position distribution for the IRTP in the intermediate regime (R2) obtained using numerical simulations with \eref{eq:Pxt_it_II}. Clearly the results show a reasonably good agreement in this regime for small values of particle mass $m$.

\subsection{Inertia dominated intermediate regime (R3)}\label{sec:ID_IR}

\begin{figure}
    \centering
    \includegraphics[width=7.5cm]{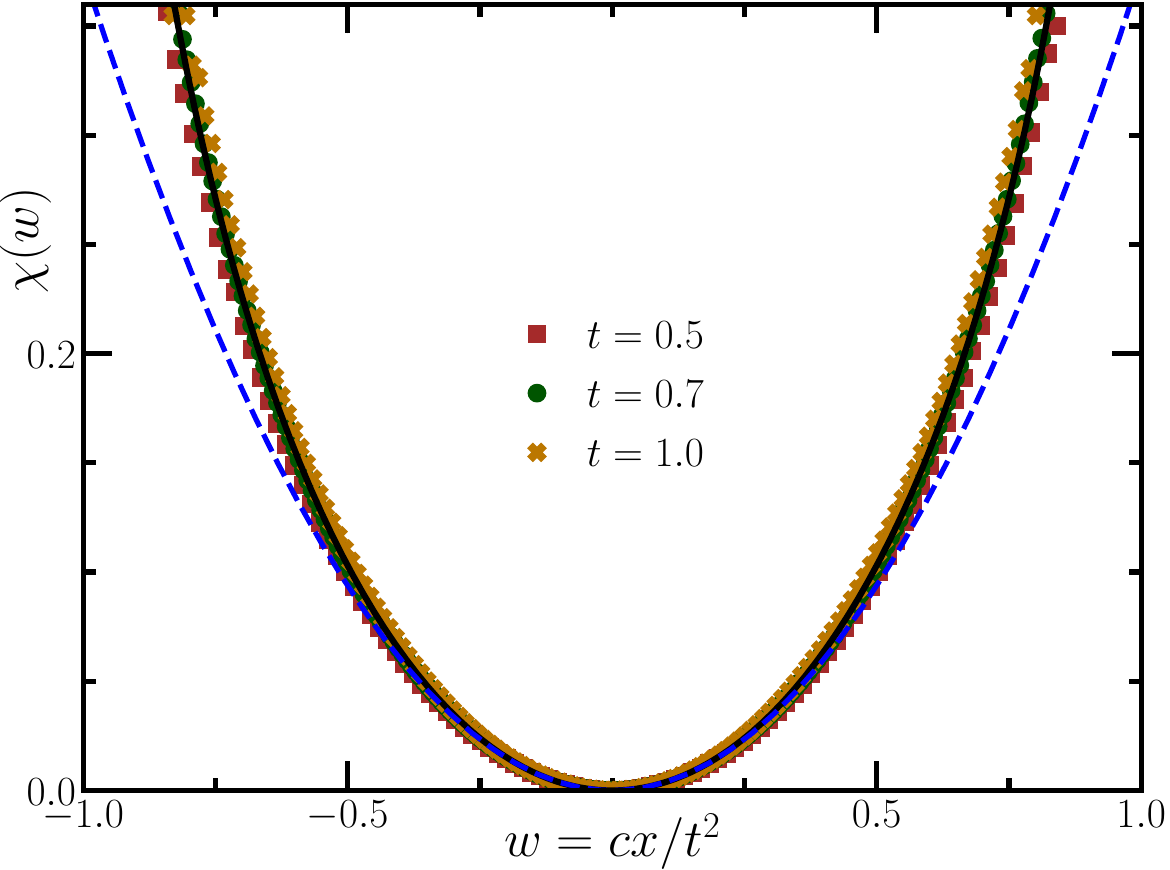}
    \caption{Inertia-dominated intermediate-time regime (R3):     
    Plot of LDF $\chi(w)$ as a function of $w=cx/t^2$ with $c=2m/a_0$, extracted from numerical simulations (symbols) using \eref{eq:chi_sim}. The solid black line corresponds to the theoretical prediction \eqref{eq:chi_z}. To illustrate the deviation from the typical Gaussian distribution, the quadratic behaviour for small $w$ is indicated by the blue dashed line. Here we have used the parameters $\tau_a=0.02, m=10.0, \gamma=1.0$ so that $\tau_m=10.0$ and $a_0=1.0$.}
    \label{fig:Px_idir}
\end{figure}

The inertia dominated intermediate-time regime emerges when the inertial time-scale is much larger compared to the active time-scale [see Fig.~\ref{fig:regimes}]. To the leading order in $\tau_a/\tau_m$,
the Langevin \erefs{eq:Le_x} and \eref{eq:Le_v} can be approximated by, 
\begin{align}
    m\ddot{x}(t) \approx a_0\sigma(t),
    \label{eq:lang-rap}
\end{align}
which describes the motion of an inertial particle subject to a stochastic dichotomous force, governed by a single time-scale $\tau_a$. The position distribution of this process in the short-time regime and the large deviation function in the late-time regime have been studied recently \cite{santra2023dichotomous, dean2021position}. It has been shown analytically that the position distribution is supported inside the light cone $x\in[-a_0 t^2/m, a_0 t^2 / m]$ and position fluctuations $x \sim t^2$ are described by the large deviation form\cite{dean2021position},
\begin{align}
    P(x,t) &\sim \exp\Big[-\frac{t}{\tau_a}\chi\big(\frac{m x}{a_0 t^2}\big)\Big],\label{eq:Pxt_ir3}
\end{align}
where, the LDF is given by,
\begin{align}
    \chi(w) = \max_{-\infty<y<\infty}\Big[-wy+1-\frac{1}{2}\sqrt{1+4y^2}-\frac{1}{4y}\sinh^{-1}(2y)\Big].\label{eq:chi_z}
\end{align}
The above equation is expected to describe the position fluctuations of the IRTP in this inertia-dominated intermediate regime. To illustrate the validity of the above prediction we extract the 
 large deviation function $\chi(w)$ from $P(x,t)$ obtained from numerical simulations following,
\bea 
\chi\left(w=\frac{m x}{a_0 t^2}\right) \asymp - \frac {\tau_a}t \log \left[\frac{P(x,t)}{P(0,t)}\right],\label{eq:chi_sim}
\eea 
which ensures $\chi(0)=0$. This is plotted in Fig.~\ref{fig:Px_idir} for different values of $t$ along with Eq.~\eqref{eq:chi_z} which shows an excellent agreement.


\subsection{Long time regime (R4)}\label{subsec:Pxt_lt}

In this section, we study the behaviour of the position distribution $P(x,t)$ of the IRTP at late times $t\gg\max(\tau_m,\tau_a)$, when the average number of tumbling events is large and the typical behaviour of the telegraphic noise $\sigma(t)$ emulates that of a white noise with strength $a_0^2\tau_a$. In this regime, the particle shows a diffusive behaviour and the typical position distribution approaches a Gaussian with an effective diffusion constant $D_\text{eff}=\tau_a a_0^2/(2\gamma^2)$. However, atypical fluctuations $x\sim a_0 t/\gamma$ are expected to carry signatures of activity giving rise to non-Gaussian tails.

For large $t$, we show $P(x,t)$ admits a large deviation form,
\begin{align}
    P(x,t) \sim \exp\Big[-\frac{t}{\tau_a}~\Phi\Big(\frac{x\gamma}{a_0 t}\Big)\Big],\label{eq:LDF}
\end{align}
where the large-deviation function (LDF)  $\Phi(z)$ is supported over the interval $z\in[-1,1]$. In the following, we compute the LDF explicitly.

Rescaling space and time, $X=\gamma^2 x / (m a_0),~T= t/\tau_m$ we rewrite the Langevin \erefs{eq:Le_x}-\eqref{eq:Le_v} in the following dimensionless form,
\begin{align}
    \dot{V} + V = \sigma(T),\label{eq:Le_LDP}
\end{align}
where, $V= dX/dT$ and $\sigma(T)$ remains a dichotomous noise which flips between $\pm 1$ with a rescaled rate $\alpha^{-1}=\tau_m/\tau_a$. 
Recall that we are interested to find $P(X,T)$, the  distribution of the position $X=\intop_{0}^{T}ds~V(s)$ of the particle at time $T$. 
The corresponding characteristic function 
is denoted by 
\begin{align}
   \mathscr{Z}(\lambda,T) & \equiv \left \langle e^{\lambda X}\right \rangle
    =\int \mathcal{D}[\sigma(s)]~\mathbb{P}[\sigma(s)] e^{\lambda X[\sigma(s)]},\label{eq:pint}
\end{align}
where $\mathbb{P}[\sigma(s)]$ represents the probability of a trajectory ${\bm \sigma}(s)=\{\sigma(s)~;0\le s\le T\}$.
For large $T \gg \alpha$, the probability of finding the particle at $X \sim T$ gets dominant contribution from the optimal trajectory that takes the particle starting from $0$ to $X$ at time $T$. 
To find this contribution we follow the formalism developed in Ref.\cite{smith2022nonequilibrium}. This formalism relies on finding an effective coarse-grained description of the Langevin Eq.~\eqref{eq:Le_LDP}
under time-scale separation $\alpha \ll 1$ {\it i.e.,} $\tau_a \ll \tau_m$.

The formal solution of \eref{eq:Le_LDP} is given by,
\begin{align}
  V(T) = \int_0^T ds \,   e^{-(T-s)}~\sigma( s). \label{eq:sol-tildev}
\end{align}
The above integral can be broken into $M$ equal intervals, $T= \sum_{i=1}^M (s_{i+1}-s_i)  = M \Delta s$ with $s_1=0$ and $s_{M+1}=T$ and  we can write,
\begin{align}
  V(T) = \sum_{i=1}^M \int_{s_i}^{s_{i+1}}d\tilde{s}_i \, e^{-(T-\tilde {s}_i)} \sigma(\tilde{s}_i). \label{eq:int_sum}
\end{align}
Since, $\sigma(\tilde{s})$ changes over a time scale $\alpha \ll 1$, choosing $\alpha \ll  \Delta s \equiv (s_{i+1}-s_i) \ll T$, one can approximate the sum in Eq.~\eqref{eq:int_sum} as, 
\begin{align}
  V(T) \approx \sum_{i=1}^M \Delta s ~e^{-(T-s_i)} \bar{\sigma}_i,\label{eq:LE-approx-1}
\end{align}
where
\begin{align}
\bar{\sigma}_i =\frac{1}{\Delta s}\int_{s_i}^{s_{i}+\Delta s}d\tilde{s} \,  \sigma(\tilde{s}). \label{eq:bar-sig}
\end{align}
Note that $\bar{\sigma}_{i}$ is now a continuous noise and is bounded in the interval $[-1,1]$. Equation \eqref{eq:LE-approx-1} is equivalent to a coarse-grained version of the Langevin equation \eqref{eq:Le_LDP},
\begin{align}
    \dot{V}(s)+V(s)=\bar{\sigma}(s).\label{eq:Le_LDPc}
\end{align}
Formally solving this equation gives
\begin{align}
X(T)=\int_{0}^{T}ds~\bar{\sigma}(s)(1-e^{s-T}). \label{eq:X_vt}
\end{align}
The noise $\bar \sigma (s)$, defined in Eq.~\eqref{eq:bar-sig}, which denotes the coarse-grained version of the dichotomous noise $\sigma(s)$, is essentially proportional to the displacement of an ordinary overdamped RTP\cite{Malakar2018}.  It has been shown that, for $\Delta s \gg \alpha$, the probability distribution of $\bar \sigma_i$ is asymptotically given by 
\begin{align}
    P(\bar{\sigma})\sim\exp\Big[-\frac{\Delta s}{\alpha}S(\bar{\sigma})\Big], \label{eq:LDP_s}
\end{align}
with the LDF 
\begin{align}
    S(w) = 1-\sqrt{1-w^2},~~\text{for}~w\in[-1,1]. \label{eq:LDF_sig}
\end{align}

\begin{figure}
    \centering
    \includegraphics[width=8.5cm]{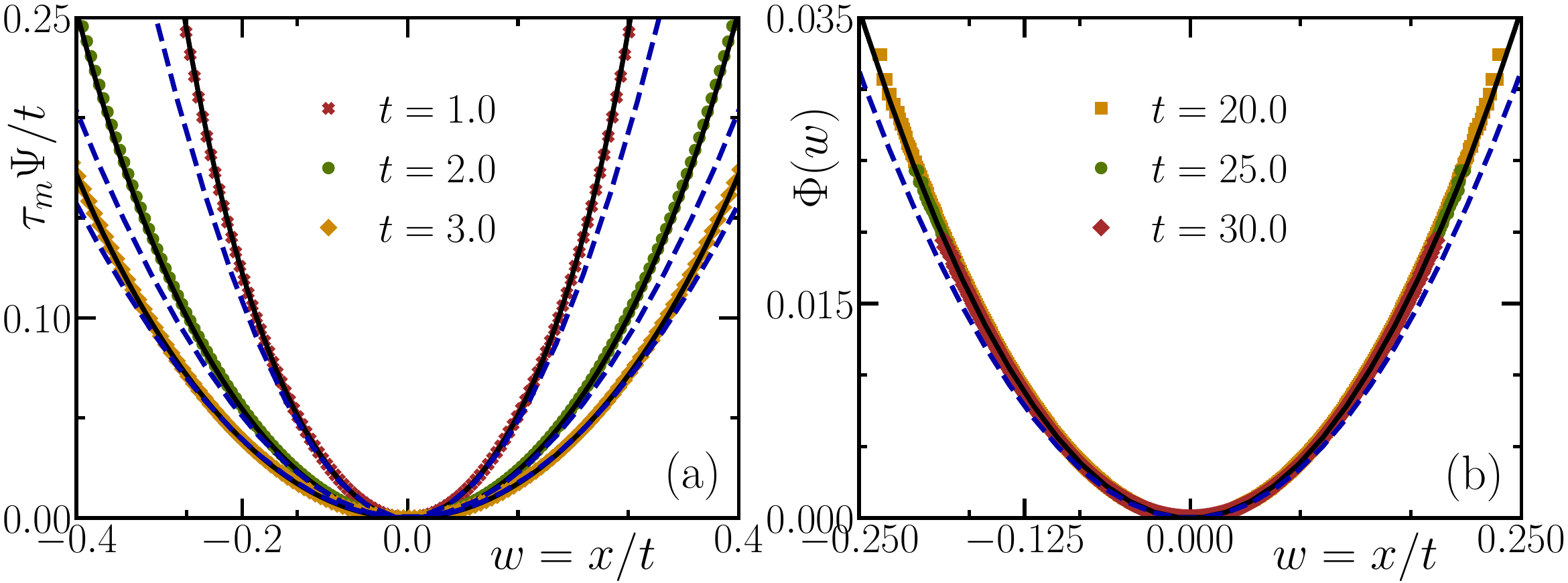}
    \caption{Long-time regime (R4): (a) Plot of $\tau_m \Psi(X,T)/t$ as a function of $x/t$ (symbols) extracted from numerical simulation data using   \eref{eq:Psi_sim}. The corresponding analytical prediction is plotted (solid black lines) using \eref{eq:PsiZ}. The typical Gaussian fluctuations are indicated by the corresponding quadratic approximation of the LDF (blue dashed lines). (b) Convergence of the LDF $\Psi(X,T)$ to the overdamped limit $T \Phi(X/T)$ (solid black line) in the $T \gg 1$ regime keeping other parameters fixed. 
    For both the plots, we have used $\tau_a=0.05, m=1.0,\gamma=1.0$  and $a_0=1.0$.}
    \label{fig:Px_LDP}
\end{figure}

To find the statistics of $X(T)$ in Eq.~\eqref{eq:X_vt} we, however, need the joint distribution $P(\bar{\sigma}_1,\dots,\bar{\sigma}_M)$.
For $\alpha/\Delta s \ll 1$, the random variables $\{\bar{\sigma}_i\}$ become statistically independent so that their joint distribution function is given by, 
\begin{align}
    P(\bar{\sigma}_1,\dots,\bar{\sigma}_N)\sim\exp\Big[-\frac{1}{\alpha}\sum_{i=1}^{M}\Delta s~S(\bar{\sigma}_i)\Big],\label{eq:LDP_sdc}
\end{align}
which, in the continuum limit, ($\Delta s \to 0,~\alpha \to 0$ with $0<\alpha/\Delta s \ll 1$) provides the probability distribution $P[\bar \sigma(s)]$ for the trajectory $\bar{\bm \sigma}(s)=\{\bar{ \sigma}(s); 0 \le s \le T \}$. The distribution $P[\bar {\bm \sigma}(s)]$ has the following large deviation form
\begin{align}
    \mathbb{P}[\bar{\bm \sigma}(s)]\sim\exp\Big[-\frac{1}{\alpha}\Sigma[\bar{\sigma}(s)]\Big],\label{eq:LDP_scc}
\end{align}
where,
\begin{align}
    \Sigma[\bar{\bm \sigma}(s)]=\intop_{0}^{T}ds~S(\bar{\sigma}(s)).
\end{align}
Using \eqref{eq:X_vt} and \eqref{eq:LDP_scc}, the characteristic function in \eref{eq:pint} can be recast in terms of the coarse grained noise $\bar{\sigma}(s)$ as,
\begin{align}
    \mathscr{Z}(\lambda,T) &=\int \mathcal{D}[\bar{\sigma}(s)]~\mathbb{P}[\bar{\sigma}(s)] e^{\lambda X[\bar{\sigma}(s)]}, \label{eq:pint-cg} \\
    &=\int \mathcal{D}[\bar{\sigma}(s)]~\exp\left[ -\frac{1}{\alpha} \int_0^Tds [S(\bar{\sigma}) - \tilde{\lambda} \bar{\sigma}(1-e^{s-T})] \right], \notag 
\end{align}
where $\tilde{\lambda}=\lambda \alpha$.

For large $\alpha \ll 1$, one can perform this path integral using saddle point method to obtain
\begin{align}
    \mathscr{Z}(\lambda,T)\sim \exp\big[-\frac{1}{\alpha}\mathscr{S}(\lambda \alpha,T)\big],\label{eq:Z-ldf}
\end{align}
where,
\begin{align}
    \mathscr{S}(\tilde{\lambda},T)=\min_{\bar{\sigma}(s)}\int_0^Tds [S(\bar{\sigma}) - \tilde{\lambda} \bar{\sigma}(1-e^{s-T})].\label{eq:saddle}
\end{align}
This minimization is performed in Appendix \ref{sec:AppC} in detail [see \erefs{eq:X_lT}].
The distribution $P(X,T)$ can be computed after performing inverse Laplace transform of $\mathscr{Z}(\lambda, T)$. The large deviation form of $\mathscr{Z}$ in Eq.~\eqref{eq:Z-ldf}, suggests a large deviation form of the position distribution
\begin{align}
P(X,T) \approx \exp \left[ -\frac{1}{\alpha} \Psi\left(X,T \right) \right], \label{eq:P_XT}
\end{align}
where $\Psi(X,T)$ can be computed via Legendre-transform
\begin{align}
\Psi(X,T) &=  \left[\mathscr{S}(\tilde{\lambda}^\ast,T) + X \tilde{\lambda}^*  \right], \label{eq:PsiZ}
\end{align}
with $\tilde{\lambda}^*$ being the solution of the equation
\begin{align}
\left. \frac{d \mathscr{S}(\tilde{\lambda},T)}{d \tilde{\lambda}}\right |_{\tilde {\lambda}^\ast} = -X. \label{eq:lstar}
\end{align}
For any given $X$, we can solve the above equation for $\tilde{\lambda}^*$. Substituting this solution in Eq.~\eqref{eq:PsiZ} and using it in \eref{eq:P_XT}, we get the distribution $P(X,T)$.

For large $T$, the LDF $\mathscr{S}(\tilde{\lambda},T)$ can be evaluated analytically to the leading order in $T$, which yields,  
\bea
\mathscr{S}(\tilde{\lambda},T) = T \mu(\tilde{\lambda})~\text{with} ~~\mu(\tilde \lambda)=1-\sqrt{1+ \tilde \lambda^2} \label{eq:LDF_cf}
\eea
This, in turn, implies that for large $T$, the distribution $P(X,T)$ has the following large deviation form,
\begin{align}
P(X,T) \approx \exp \left[ -\frac{T}{\alpha} \Phi\left(\frac{X}{T} \right) \right]. \label{eq:phi_def}
\end{align}
with the LDF, 
\begin{align}
    \Phi(w) &= \max_{-\infty<\tilde \lambda<\infty}\big[\mu(\tilde \lambda)+w \tilde \lambda\big],\cr
            &=1-\sqrt{1-w^2}.\label{eq:LDF_ORTP}
\end{align}
Going back to the unscaled variables $(x,t)$, the above expression leads to the large deviation form for $P(x,t)$ mentioned in Eq.~\eqref{eq:LDF}. Note that the LDF quoted in \eref{eq:LDF_ORTP} is same as the LDF of an ordinary overdamped RTP, which is indeed expected in the $t \gg \tau_m$ limit.

To verify the above predictions, we extract the function $\Psi(X,T)$ from $P(x,t)$, obtained from numerical simulations following
\begin{align}
    \Psi\left(X=\frac{\gamma^2 x}{m a_0},T=\frac{t}{\tau_m}\right) \asymp- \frac {\tau_a}{\tau_m} \log \left[\frac{P(x,t)}{P(0,t)}\right], \label{eq:Psi_sim}
\end{align}
which ensures that $\Psi(0,T)=0$. Figure.~\ref{fig:Px_LDP}(a) shows the plot for $\Psi(X,T)/ T $ as a function of $x/t$ for different values of $t$. We make this particular choice of scale for plotting since for large $T$ we expect $\Psi(X,T) \to T\Phi(X/T)$. For $t \sim O(\tau_m)$, we observe a very good agreement with the theoretical result obtained from \eref{eq:PsiZ}. 
The convergence of $\Psi(X,T)$ to the LDF $\Phi(X/T)$ is illustrated in Fig.~\ref{fig:Px_LDP}(b). The excellent data collapse validates our prediction of the LDF.

As a final remark, we mention that, from the large deviation form \eqref{eq:LDF_ORTP} one can indeed show that the kurtosis $\kappa(t)$ decays as $1/t$ for large $t$, as obtained from the exact computation of the moments [see Fig.~\ref{fig:kurtosis}].

\section{First-Passage Properties}\label{sec:Qt}
Persistent nature of active motion leads to interesting first passage properties in the presence of absorbing boundaries\cite{Malakar2018,santra2021active,basu2018active}. In this section, we explore the first-passage properties of the IRTP across various dynamical regimes. The first-passage property is most conveniently characterized by the survival probability $\mathscr{Q}(t)$ defined as the probability that the particle starting at some initial position $x_0\ge 0$, does not cross the origin upto time $t$. In the following, we separately investigate the behaviour of the survival probability for each of the four distinct regimes.

At short-times (R1), when $t\ll\min(\tau_m,\tau_a)$ the particle moves deterministically surviving with probability $\sim 1$ for initial propulsion direction $\sigma=+1$. On the other hand in the late-time regime R4, when $t\gg\max(\tau_m,\tau_a)$, the particle behaves diffusively and we expect a decay $\sim t^{-1/2}$ at late times. 

\begin{figure}
    \centering
    \includegraphics[width=7.5cm]{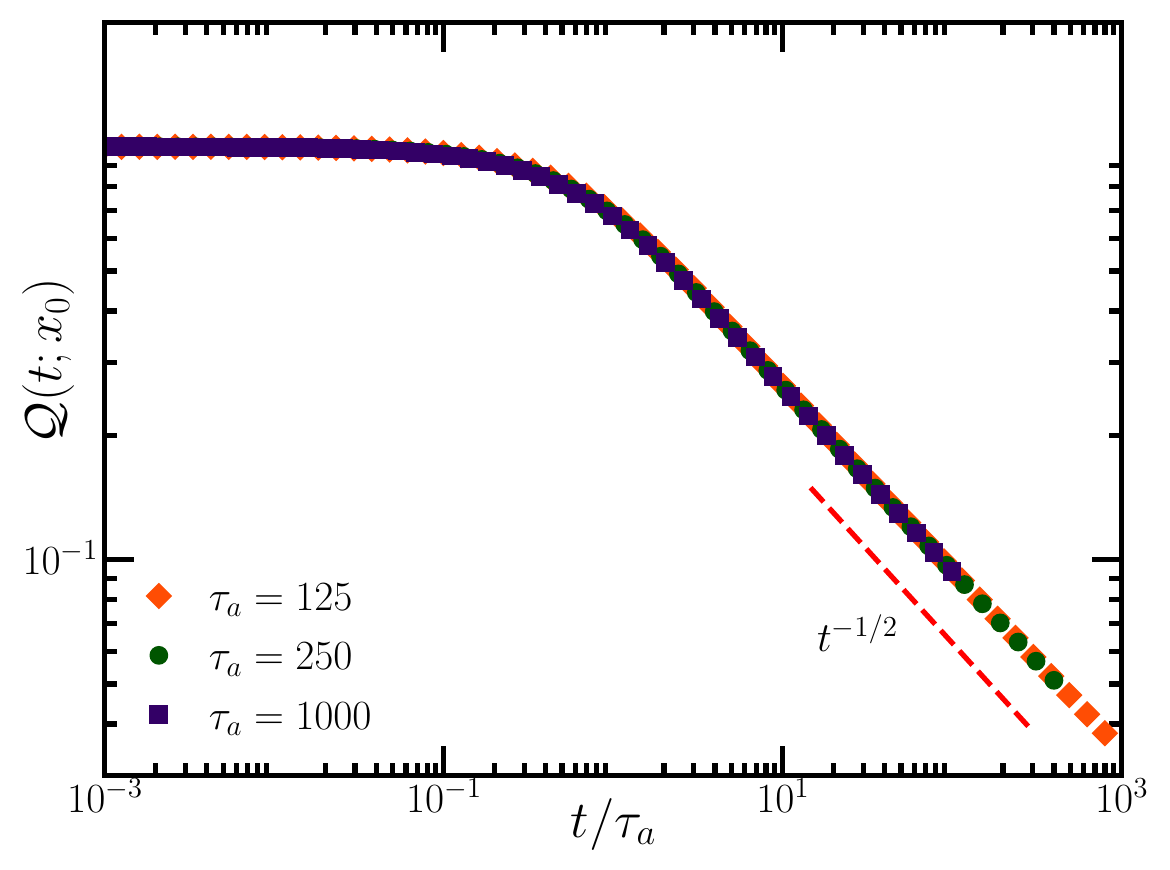}
    \caption{Survival probability in the activity-dominated case ($\tau_m\ll\tau_a$): Plot of $\mathscr{Q}(t)$ as a function of $t/\tau_a$  for different values of $\tau_a$. The red dashed line  indicates the  $t^{-1/2}$ decay in the $t \gg \tau_a$ regime [see \eref{eq:AD_IR_S_scaling}]. The data are obtained from numerical simulations using $m=5.0, \gamma=1.0$ so that $\tau_m=5.0$ and $a_0=1.0$.}
    \label{fig:surv-ta-gg-tm}
\end{figure}

Interestingly, the behaviour of the survival probability $\mathscr{Q}(t)$ in the intermediate-regimes depends on the ordering of the time-scales $\tau_m$ and $\tau_a$. In the activity-dominated intermediate regime R2, $\tau_m \ll t\ll\tau_a$, the IRTP effectively undergoes a damped deterministic motion. Consequently, it  can still survive with probability $\simeq 1$ starting with $\sigma=+1$. Now moving from R2 to R4, keeping $\tau_m\ll\tau_a$, we expect a crossover behaviour of the form\cite{basu2018active},
\begin{align}
    \mathscr{Q}(t) = \mathcal{F}_{a}(t/\tau_a),\label{eq:AD_IR_S_scaling}
\end{align}
where the crossover function has the asymptotic behaviour,
\begin{align}
\mathcal{F}_{a}(u) \sim
    \begin{cases}
        1 &~\text{for}~u\ll 1 \\
        u^{-1/2} &~\text{for}~u\gg 1
    \end{cases}
\end{align}
This scaling behaviour is numerically demonstrated in Fig.~\ref{fig:surv-ta-gg-tm} for three values of $\tau_a$ all much larger than $\tau_m=5.0$. The excellent data collapse verifies our prediction in \eref{eq:AD_IR_S_scaling}.

The scenario however is very different in the inertia-dominated intermediate case $\tau_a \ll \tau_m$. As already discussed in Sec.~\ref{sec:ID_IR}, in this case  the dynamics of the particle in regime $R3$ can be effectively described by the Langevin equation (\ref{eq:lang-rap}). For $t \gg \tau_a$, this equation effectively reduces to a random acceleration process.  Survival probability for this process 
is known exactly and shows an anomalous $t^{-1/4}$ decay \cite{burkhardt2007random,majumdar2010time}. On the other hand, for $t\gg \tau_m$ the typical motion of the particle becomes diffusive [as also seen from the moment in \eref{eq:x2_D}] where $\mathscr{Q}(t) \sim t^{-1/2}$ is expected\cite{marshall1985drop,burkhardt2014first}. This suggests a crossover from $\mathscr{Q}(t) \sim t^{-1/4}$ to $\mathscr{Q}(t) \sim t^{-1/2}$ as $t$ crosses the time-scale $\tau_m$. 

\begin{figure}
    \centering
    \includegraphics[width=8.5cm]{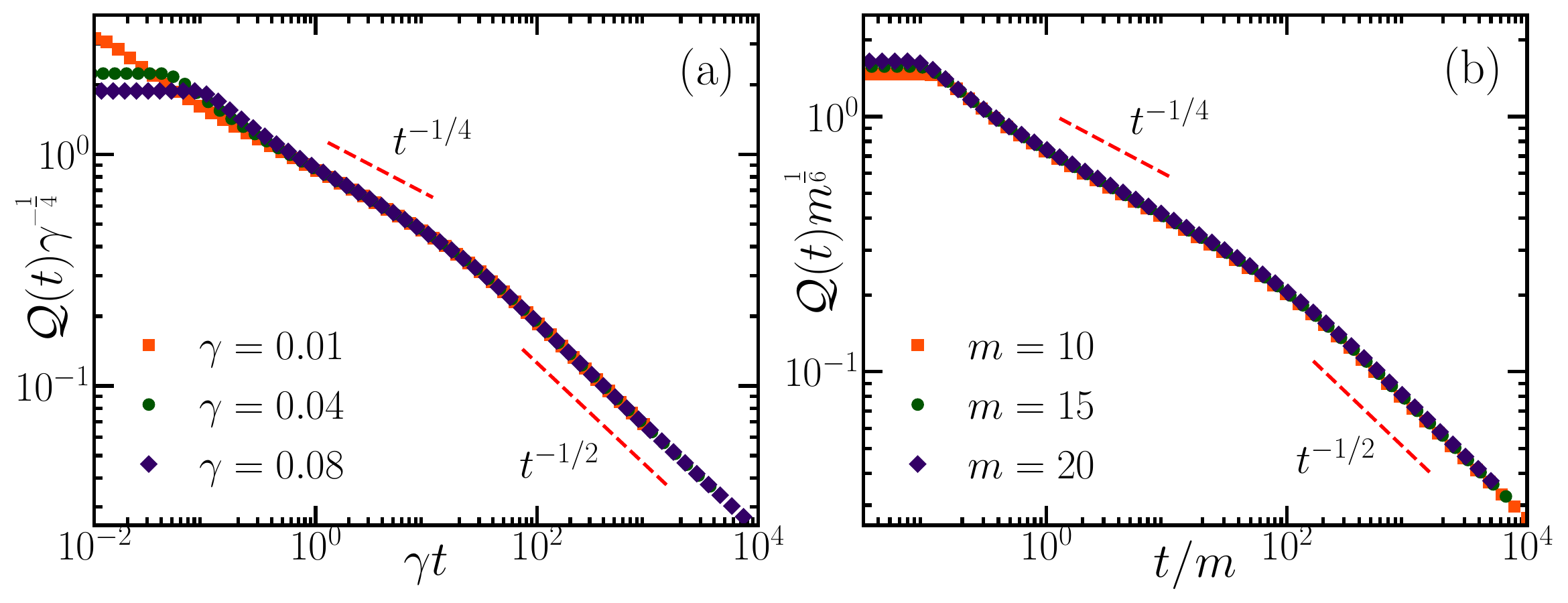}
    \caption{Time-evolution of the survival probability in the inertia-dominated case ($\tau_a\ll\tau_m$) for (a) fixed $m$ and (b) fixed $\gamma$.  Both the plots show  a crossover from $t^{-1/4}$ to $t^{-1/2}$. Panel (a) verifies the $\gamma^{-1/4}$ dependence in \eref{eq:Fm_long_t} whereas figure (b) verifies the $m^{1/12}$ dependence in \eref{eq:Fm_short_t}. Here we have used parameter values $m=10.0, \tau_a=0.2$ and $a_0=1.0$ for (a), and $\gamma=0.02, \tau_a=0.2$ and $a_0=1.0$ for (b).}
    \label{fig:St}
\end{figure}

Using data from numerical simulations, we illustrate this crossover behaviour in Fig.~\ref{fig:St}(a) for fixed $m$ and in Fig.~\ref{fig:St}(b) for fixed $\gamma$. The excellent data collapse observed in these figures for the particular choices of the scaled variables indicates the following scaling form for the survival probability in the inertia dominated case ($\tau_a\ll\tau_m$), 
\begin{align}
    \mathscr{Q}(t) = C \left(\frac{x_0 \tau_a \gamma}{a_0 \tau_m^2}  \right)^{1/12} {\cal F}_m(t /\tau_m),\label{eq:ID_IR_S_scaling}
\end{align}
where, $\tau_m=m/\gamma$, and $C$ is some dimensionless number, independent of $m$ and $\gamma$.  The scaling function has the asymptotic behaviour, 
\begin{align}
\mathcal{F}_{m}(u) \sim
    \begin{cases}
        u^{-1/4} &~\text{for}~u \ll 1 \\
        u^{-1/2} &~\text{for}~u\gg 1
    \end{cases}
\end{align}
Note that, the above scaling form implies that, for $t \ll \tau_m$, 
\bea
\mathscr{Q}(t) \sim m^{1/12} t^{-1/4},
\label{eq:Fm_short_t}
\eea
whereas, at late-times $t \gg \tau_m$, , we have,
\bea
\mathscr{Q}(t) \sim m^{1/3} \gamma^{-1/4} t^{-1/2},
\label{eq:Fm_long_t}
\eea 

\section{Conclusion}\label{sec:conc}
In this work, we have studied the dynamical behaviour of an inertial Run-and-Tumble particle moving on a line. We show that, such an inertial active particle exhibits a much richer dynamical behaviour, in contrast to a passive inertial particle, which shows a crossover from a ballistic motion at short-times to a diffusive one at late-times. The presence of both inertia and activity gives rise to two distinct time-scales in the dynamics. The interplay between these time-scales leads to the emergence of four different dynamical regimes [see Fig.~\ref{fig:regimes}] where the particle displays different statistical properties.

To illustrate the difference in the dynamical behaviour of the particle in these regimes, we first compute the moments and correlation of the position and velocity. We find that the MSD of the position grows as $\sim t^\theta$ with $\theta$ taking different values in the different regimes. In particular, $\theta=4$ in the short-time regime, indicating a ballistic behaviour, which crosses over to a diffusive one ($\theta=1$) via an intermediate regime, where $\theta=2$ or $\theta=3$ depending on the relative strength of the two time-scales [see Table~\ref{table:mom_sct}].  

We also compute the position distribution analytically in the different regimes when the time-scales are well separated.  The ballistic behaviour at short-time regime leads to a $U$-shaped position distribution, with divergences near $x \sim t^2$ implying that the particle is most likely to be away from its initial position. A similar feature is obtained in the activity dominated intermediate regime, while the inertia dominated intermediate regime shows a distribution peaked at the origin. We analytically compute the distributions in each of these regimes using simplified effective dynamics. At late-times, the typical position fluctuations are Gaussian, with signature of activity manifest at the tails, which satisfies a large deviation form; we compute the large deviation function analytically.  

We also study the first-passage properties of the position of the IRTP and show that the dynamical regimes are characterized by distinct persistence exponents. Using theoretical arguments, we predict the values of these exponents. Moreover, we show that, the survival probability satisfies two very different scaling forms in the activity dominated and inertia dominated scenarios. 

A natural open question is to understand the crossover of 
the distributions across different regimes (R1 $\to$ R4), as the time-scales are tuned. Extending the study of IRTPs to higher dimensions would also be interesting, particularly for experimental applications. IRTP provides a simple set-up to investigate how inertia affects the collective behaviour of active particles. In particular, it would be intriguing to investigate the statistical properties of a tagged particle in a bath of IRTPs. Finally, the role of inertia could be further examined in relation to more complex active particle systems, such as the direction-reversing active Brownian particle\cite{santra2021active}.

\section*{Acknowledgement}
A. K. would like to acknowledge the support of DST, Government of India Grant under Project No. CRG/2021/002455 and the MATRICS grant MTR/2021/000350 from Anusandhan National Research Foundation (ANRF), Government of India. A. K. would also like to acknowledge the Department of Atomic Energy, Government of India, for their support under Project No. RTI4001. U. B. acknowledges support from Anusandhan National Research Foundation (ANRF), Government of India under the MATRICS scheme (No. MTR/2023/000392).

\section*{Author Declarations} 
\noindent {\bf Conflict of Interest}

The authors have no conflicts to disclose.

\section*{Data Availability}

The data generated through numerical simulations are presented in the figures, there are no additional data.

\appendix
\section{Formal solution of Langevin equation}\label{sec:AppB}
To solve the Langevin \erefs{eq:Le_x} and \eqref{eq:Le_v}, we first write down the general solution for the velocity $v(t)$, given by,
\begin{align}
    v(t) &= v_0e^{-t/\tau_m}+\frac{a_0}{m}\int_{0}^{t}ds~e^{-(t-s)/\tau_m}\sigma(s),\label{eq:vt_f}
\end{align}
where $v(0)=v_0$.
It is now straightforward to write down the  solution for the particle position first by substituting \eref{eq:vt_f} into \eref{eq:Le_x}, 
\begin{align}
\begin{split}
    x(t) &= x_0 + v_0 \tau_m \big(1-e^{-t/\tau_m}\big) \\
    &~~+ \frac{a_0}{\gamma}\int_{0}^{t}ds~\big[1-e^{-(t-s)/\tau_m}\big]\sigma(s),
    \end{split}
    \label{eq:xt_f}
\end{align}
where $x(0)=x_0$. For the initial condition $x_0=v_0=0$, the second-order correlations can be computed explicitly using \eqref{eq:vt_f} and \eqref{eq:xt_f} and the autocorrelation function for the dichotomous noise\cite{haunggi1994colored},
\begin{align}
    \la \sigma(t_1)\sigma(t_2) \ra = e^{-2 |t_1 - t_2|/\tau_a},
\end{align}
resulting in expressions quoted in \erefs{eq:M_02}-\eqref{eq:M_20}. However, the computation of higher moments becomes increasingly cumbersome using this method. Instead, we resort to the recursive procedure discussed in Sec.~\ref{sec:moments} for the computation of the higher order moments such as kurtosis.


\section{Detailed computation of the large deviation function} \label{sec:AppC}In this Appendix we provide the details of the calculation leading to the LDFs $\Psi(X,T)$ and $\Phi(w)$ defined in \erefs{eq:P_XT} and \eqref{eq:phi_def}.
We begin by performing the minimization in \eqref{eq:saddle}, which leads to the Euler-Lagrange equation,
\begin{align}
    \frac{d}{ds}\Big[\frac{\partial\mathcal{L}(\bar{\sigma},\dot{\bar{\sigma}},s)}{\partial \dot{\bar{\sigma}}}\Big]=\frac{\partial\mathcal{L}(\bar{\sigma},\dot{\bar{\sigma}},s)}{\partial \bar{\sigma}},\label{eq:EL}
\end{align}
with the Lagrangian,
\begin{align}
    \mathcal{L}(\bar{\sigma},\dot{\bar{\sigma}},s) = S(\bar{\sigma}) - \tilde{\lambda}\bar{\sigma}(1-e^{s-T}),\label{eq:Lag}
\end{align}
where $S(w)=1-\sqrt{1-w^2}$.
Substituting  \eref{eq:Lag} into \eref{eq:EL}, we get,
\begin{align}
    \frac{\bar{\sigma}(s)}{\sqrt{1-\bar{\sigma}^{2}(s)}} = \tilde{\lambda}(1-e^{s-T}). \label{eq:EL_Lag}
\end{align}
Solving \eref{eq:EL_Lag} for $\bar{\sigma}(s)$ gives the optimal trajectory of the coarse-grained noise,
\begin{align}
    \bar{\sigma}(s)=\frac{\tilde{\lambda}(1-e^{s-T})}{1+\tilde{\lambda}^{2}(1-e^{s-T})^{2}}.\label{eq:sig_opt}
\end{align}
The position distribution can now be computed by using this optimal trajectory \eref{eq:sig_opt}. First, we substitute the solution in\eref{eq:sig_opt} in \eref{eq:saddle}, which gives,
\begin{align}
    \mathscr{S}(\tilde{\lambda},T)=\int_{0}^{T}ds~\Big[1-\sqrt{1+\tilde{\lambda}^2(1-e^{s-T})^2}\Big]. \label{eq:S_tilde-app}
\end{align}
Next, we use \eref{eq:S_tilde-app} in \eref{eq:lstar}, and get,
\begin{widetext}
\begin{align}
    X &= \frac{2 \tilde{\lambda}}{\sqrt{\tilde{\lambda} ^2+1}} \Bigg[\coth ^{-1}\Bigg(\frac{e^T\sqrt{\tilde{\lambda} ^2+1}}{\tilde{\lambda} +e^T \sqrt{\tilde{\lambda}^2(1 - e^{-T})^2+1}}\Bigg)-\tanh ^{-1}\Bigg(\frac{\tilde{\lambda} +1}{\sqrt{\tilde{\lambda} ^2+1}}\Bigg)\Bigg]+\sinh ^{-1}\big[\tilde{\lambda}(e^{-T}-1)\big]\cr
    &+\frac{1}{\tilde{\lambda}}\Big[1-\sqrt{\tilde{\lambda} ^2 e^{-2 T}(e^T-1)^2+1}\Big]\Bigg |_{\tilde {\lambda}=\tilde{\lambda}^\ast}.\label{eq:X_lT}
\end{align}
\end{widetext}
Finally one needs to solve this equation for $\tilde{\lambda}^*$ for a given $X$ and substitute it Eq.~\eref{eq:PsiZ} to obtain the desired LDF $\Psi(X,T)$.

In the late-time regime $T\gg 1$, \eref{eq:X_lT} takes the simple form,
\begin{align}
    X &= \left . \frac{\tilde{\lambda} T}{\sqrt{1+{\tilde{\lambda}^2}}}\right |_{\tilde{\lambda}^{\ast}},\label{eq:X_lT_long}
\end{align}
which can be solved explicitly to obtain,
\bea 
{\tilde{\lambda}^{\ast}} = \frac{X/T}{\sqrt{1-(X/T)^{2}}}. 
\eea 
Substituting the above equation  in \eref{eq:PsiZ}, we recover the LDF for the position distribution quoted in \eref{eq:LDF_ORTP},
\begin{align}
    \Psi(X,T) &= T\Big[1-\sqrt{1-(X/T)^2}\Big]=T\Phi(X/T).
\end{align}

\section*{References} 

\bibliographystyle{apsrev4-1}
\bibliography{cite}

\end{document}